\documentclass[%
 aps,
 reprint,
superscriptaddress,
nofootinbib,
 amsmath,amssymb,
 prd,
showpacs
]{revtex4-2}
\usepackage{graphicx}

\usepackage{aas_macros}
\usepackage{lineno}
\usepackage{booktabs}
\usepackage{hyperref}
\usepackage{multirow}
\usepackage[normalem]{ulem}
\usepackage{xcolor}
\usepackage{orcidlink}
\newcommand{\de}{\operatorname{d}}
\newcommand{\HOfromTOandSNeBF}{65.2^{+4.9}_{-6.2}\;\mathrm{km/s/Mpc}}
\newcommand{\HOfromTOandBAOBF}{74.0^{+7.2}_{-3.5}\;\mathrm{km/s/Mpc}}
\newcommand{\kmsMpc}{\,{\rm km\,s^{-1}\,Mpc^{-1}}}

\definecolor{ForestGreen}{HTML}{009B55}
\definecolor{Purple}{HTML}{99479B}
\definecolor{Gold}{HTML}{999B47}
\definecolor{Red}{HTML}{9B0055}
\definecolor{Turquiose}{HTML}{40E0D0}

\definecolor{amethyst}{rgb}{0.6, 0.4, 0.8}

\usepackage{CJK}
\newcommand{\mathhangeulke}{\text{\it\begin{CJK}{UTF8}{mj}케\end{CJK}}}
\newcommand{\mathhangeulkyu}{\text{\it\begin{CJK}{UTF8}{mj}큐\end{CJK}}}

\begin{document}

\preprint{APS/123-QED}
\title{Model-Independent Measurement of the Matter-Radiation Equality Scale in DESI 2024}

% Collaborations

%% [A] If main author
%% \collaboration{\includegraphics[height=17mm]{collabroation-logo}\\begin{equation}6pt]
%%  XXX collaboration}

%% or
%% [B] If "on behalf of"
%% \collaboration[c]{on behalf of XXX collaboration}

% \input{DESI-2025-0423_author_list_5th_april}
% Author list file generated with: mkauthlist.py 1.3.0+44.gbe076f7.dirty 
% mkauthlist.py -f --sort-firsttier --orcid -j revtex data/DESI-2024-0423_author_list.csv data/DESI-2024-0423_author_list.tex 
%% Orcid numbers may need \usepackage{orcidlink}.
%% Use \input to call the file

\author{B.~Bahr-Kalus\orcidlink{0000-0002-4578-4019}}
\email{benedict.bahrkalus@inaf.it}
\affiliation{INAF, Osservatorio Astrofisico di Torino, Via Osservatorio 20, 10025 Pino Torinese, Italy}%
\affiliation{Dipartimento di Fisica, Universit\`a degli Studi di Torino, Via P.\ Giuria 1, 10125 Torino, Italy}%
\affiliation{INFN, Sezione di Torino, Via P.\ Giuria 1, 10125 Torino, Italy}%
\affiliation{Korea Astronomy and Space Science Institute, 776, Daedeokdae-ro, Yuseong-gu, Daejeon 34055, Republic of Korea}

\author{D.~Parkinson\orcidlink{0000-0002-7464-2351}}
\email{davidparkinson@kasi.re.kr}
\affiliation{Korea Astronomy and Space Science Institute, 776, Daedeokdae-ro, Yuseong-gu, Daejeon 34055, Republic of Korea}
\affiliation{University of Science and Technology, 217 Gajeong-ro, Yuseong-gu, Daejeon 34113, Republic of Korea}

\author{K.~Lodha\orcidlink{0009-0004-2558-5655}}
\affiliation{Korea Astronomy and Space Science Institute, 776, Daedeokdae-ro, Yuseong-gu, Daejeon 34055, Republic of Korea}
\affiliation{University of Science and Technology, 217 Gajeong-ro, Yuseong-gu, Daejeon 34113, Republic of Korea}

\author{E.~Mueller}
\affiliation{Department of Physics and Astronomy, University of Sussex, Brighton BN1 9QH, U.K}

\author{E.~Chaussidon\orcidlink{0000-0001-8996-4874}}
\affiliation{Lawrence Berkeley National Laboratory, 1 Cyclotron Road, Berkeley, CA 94720, USA}

\author{A.~de~Mattia\orcidlink{0000-0003-0920-2947}}
\affiliation{IRFU, CEA, Universit\'{e} Paris-Saclay, F-91191 Gif-sur-Yvette, France}

\author{D.~Forero-Sánchez\orcidlink{0000-0001-5957-332X}}
\affiliation{Institute of Physics, Laboratory of Astrophysics, \'{E}cole Polytechnique F\'{e}d\'{e}rale de Lausanne (EPFL), Observatoire de Sauverny, Chemin Pegasi 51, CH-1290 Versoix, Switzerland}

\author{J.~Aguilar}
\affiliation{Lawrence Berkeley National Laboratory, 1 Cyclotron Road, Berkeley, CA 94720, USA}

\author{S.~Ahlen\orcidlink{0000-0001-6098-7247}}
\affiliation{Department of Physics, Boston University, 590 Commonwealth Avenue, Boston, MA 02215 USA}

\author{D.~Bianchi\orcidlink{0000-0001-9712-0006}}
\affiliation{Dipartimento di Fisica ``Aldo Pontremoli'', Universit\`a degli Studi di Milano, Via Celoria 16, I-20133 Milano, Italy}
\affiliation{INAF-Osservatorio Astronomico di Brera, Via Brera 28, 20122 Milano, Italy}

\author{D.~Brooks}
\affiliation{Department of Physics \& Astronomy, University College London, Gower Street, London, WC1E 6BT, UK}

\author{T.~Claybaugh}
\affiliation{Lawrence Berkeley National Laboratory, 1 Cyclotron Road, Berkeley, CA 94720, USA}

\author{A.~Cuceu\orcidlink{0000-0002-2169-0595}}
\affiliation{Lawrence Berkeley National Laboratory, 1 Cyclotron Road, Berkeley, CA 94720, USA}

\author{A.~de la Macorra\orcidlink{0000-0002-1769-1640}}
\affiliation{Instituto de F\'{\i}sica, Universidad Nacional Aut\'{o}noma de M\'{e}xico,  Circuito de la Investigaci\'{o}n Cient\'{\i}fica, Ciudad Universitaria, Cd. de M\'{e}xico  C.~P.~04510,  M\'{e}xico}

\author{P.~Doel}
\affiliation{Department of Physics \& Astronomy, University College London, Gower Street, London, WC1E 6BT, UK}

\author{A.~Font-Ribera\orcidlink{0000-0002-3033-7312}}
\affiliation{Institut de F\'{i}sica d’Altes Energies (IFAE), The Barcelona Institute of Science and Technology, Edifici Cn, Campus UAB, 08193, Bellaterra (Barcelona), Spain}

\author{E.~Gaztañaga}
\affiliation{Institut d'Estudis Espacials de Catalunya (IEEC), c/ Esteve Terradas 1, Edifici RDIT, Campus PMT-UPC, 08860 Castelldefels, Spain}
\affiliation{Institute of Cosmology and Gravitation, University of Portsmouth, Dennis Sciama Building, Portsmouth, PO1 3FX, UK}
\affiliation{Institute of Space Sciences, ICE-CSIC, Campus UAB, Carrer de Can Magrans s/n, 08913 Bellaterra, Barcelona, Spain}

\author{S.~Gontcho A Gontcho\orcidlink{0000-0003-3142-233X}}
\affiliation{Lawrence Berkeley National Laboratory, 1 Cyclotron Road, Berkeley, CA 94720, USA}

\author{G.~Gutierrez}
\affiliation{Fermi National Accelerator Laboratory, PO Box 500, Batavia, IL 60510, USA}

\author{K.~Honscheid\orcidlink{0000-0002-6550-2023}}
\affiliation{Center for Cosmology and AstroParticle Physics, The Ohio State University, 191 West Woodruff Avenue, Columbus, OH 43210, USA}
\affiliation{Department of Physics, The Ohio State University, 191 West Woodruff Avenue, Columbus, OH 43210, USA}
\affiliation{The Ohio State University, Columbus, 43210 OH, USA}

\author{D.~Huterer\orcidlink{0000-0001-6558-0112}}
\affiliation{Department of Physics, University of Michigan, 450 Church Street, Ann Arbor, MI 48109, USA}
\affiliation{University of Michigan, 500 S. State Street, Ann Arbor, MI 48109, USA}

\author{M.~Ishak\orcidlink{0000-0002-6024-466X}}
\affiliation{Department of Physics, The University of Texas at Dallas, 800 W. Campbell Rd., Richardson, TX 75080, USA}

\author{R.~Kehoe}
\affiliation{Department of Physics, Southern Methodist University, 3215 Daniel Avenue, Dallas, TX 75275, USA}

\author{S.~Kent\orcidlink{0000-0003-4207-7420}}
\affiliation{Fermi National Accelerator Laboratory, PO Box 500, Batavia, IL 60510, USA}
\affiliation{Department of Astronomy and Astrophysics, University of Chicago, 5640 South Ellis Avenue, Chicago, IL 60637, USA}

\author{D.~Kirkby\orcidlink{0000-0002-8828-5463}}
\affiliation{Department of Physics and Astronomy, University of California, Irvine, 92697, USA}

\author{T.~Kisner\orcidlink{0000-0003-3510-7134}}
\affiliation{Lawrence Berkeley National Laboratory, 1 Cyclotron Road, Berkeley, CA 94720, USA}

\author{A.~Kremin\orcidlink{0000-0001-6356-7424}}
\affiliation{Lawrence Berkeley National Laboratory, 1 Cyclotron Road, Berkeley, CA 94720, USA}

\author{O.~Lahav}
\affiliation{Department of Physics \& Astronomy, University College London, Gower Street, London, WC1E 6BT, UK}

\author{M.~Landriau\orcidlink{0000-0003-1838-8528}}
\affiliation{Lawrence Berkeley National Laboratory, 1 Cyclotron Road, Berkeley, CA 94720, USA}

\author{L.~Le~Guillou\orcidlink{0000-0001-7178-8868}}
\affiliation{Sorbonne Universit\'{e}, CNRS/IN2P3, Laboratoire de Physique Nucl\'{e}aire et de Hautes Energies (LPNHE), FR-75005 Paris, France}

\author{C.~Magneville}
\affiliation{IRFU, CEA, Universit\'{e} Paris-Saclay, F-91191 Gif-sur-Yvette, France}

\author{M.~Manera\orcidlink{0000-0003-4962-8934}}
\affiliation{Institut de F\'{i}sica d’Altes Energies (IFAE), The Barcelona Institute of Science and Technology, Edifici Cn, Campus UAB, 08193, Bellaterra (Barcelona), Spain}
\affiliation{Departament de F\'{i}sica, Serra H\'{u}nter, Universitat Aut\`{o}noma de Barcelona, 08193 Bellaterra (Barcelona), Spain}

\author{P.~Martini\orcidlink{0000-0002-4279-4182}}
\affiliation{Center for Cosmology and AstroParticle Physics, The Ohio State University, 191 West Woodruff Avenue, Columbus, OH 43210, USA}
\affiliation{The Ohio State University, Columbus, 43210 OH, USA}
\affiliation{Department of Astronomy, The Ohio State University, 4055 McPherson Laboratory, 140 W 18th Avenue, Columbus, OH 43210, USA}

\author{A.~Meisner\orcidlink{0000-0002-1125-7384}}
\affiliation{NSF NOIRLab, 950 N. Cherry Ave., Tucson, AZ 85719, USA}

\author{R.~Miquel}
\affiliation{Institut de F\'{i}sica d’Altes Energies (IFAE), The Barcelona Institute of Science and Technology, Edifici Cn, Campus UAB, 08193, Bellaterra (Barcelona), Spain}
\affiliation{Instituci\'{o} Catalana de Recerca i Estudis Avan\c{c}ats, Passeig de Llu\'{\i}s Companys, 23, 08010 Barcelona, Spain}

\author{J.~Moustakas\orcidlink{0000-0002-2733-4559}}
\affiliation{Department of Physics and Astronomy, Siena College, 515 Loudon Road, Loudonville, NY 12211, USA}

\author{S.~Nadathur\orcidlink{0000-0001-9070-3102}}
\affiliation{Institute of Cosmology and Gravitation, University of Portsmouth, Dennis Sciama Building, Portsmouth, PO1 3FX, UK}

\author{N.~Palanque-Delabrouille\orcidlink{0000-0003-3188-784X}}
\affiliation{IRFU, CEA, Universit\'{e} Paris-Saclay, F-91191 Gif-sur-Yvette, France}
\affiliation{Lawrence Berkeley National Laboratory, 1 Cyclotron Road, Berkeley, CA 94720, USA}

\author{W.~J.~Percival\orcidlink{0000-0002-0644-5727}}
\affiliation{Department of Physics and Astronomy, University of Waterloo, 200 University Ave W, Waterloo, ON N2L 3G1, Canada}
\affiliation{Perimeter Institute for Theoretical Physics, 31 Caroline St. North, Waterloo, ON N2L 2Y5, Canada}
\affiliation{Waterloo Centre for Astrophysics, University of Waterloo, 200 University Ave W, Waterloo, ON N2L 3G1, Canada}

\author{F.~Prada\orcidlink{0000-0001-7145-8674}}
\affiliation{Instituto de Astrof\'{i}sica de Andaluc\'{i}a (CSIC), Glorieta de la Astronom\'{i}a, s/n, E-18008 Granada, Spain}

\author{I.~P\'erez-R\`afols\orcidlink{0000-0001-6979-0125}}
\affiliation{Departament de F\'isica, EEBE, Universitat Polit\`ecnica de Catalunya, c/Eduard Maristany 10, 08930 Barcelona, Spain}

\author{A.~J.~Ross\orcidlink{0000-0002-7522-9083}}
\affiliation{Center for Cosmology and AstroParticle Physics, The Ohio State University, 191 West Woodruff Avenue, Columbus, OH 43210, USA}
\affiliation{Department of Astronomy, The Ohio State University, 4055 McPherson Laboratory, 140 W 18th Avenue, Columbus, OH 43210, USA}
\affiliation{The Ohio State University, Columbus, 43210 OH, USA}

\author{G.~Rossi}
\affiliation{Department of Physics and Astronomy, Sejong University, 209 Neungdong-ro, Gwangjin-gu, Seoul 05006, Republic of Korea}

\author{L.~Samushia\orcidlink{0000-0002-1609-5687}}
\affiliation{Abastumani Astrophysical Observatory, Tbilisi, GE-0179, Georgia}
\affiliation{Department of Physics, Kansas State University, 116 Cardwell Hall, Manhattan, KS 66506, USA}
\affiliation{Faculty of Natural Sciences and Medicine, Ilia State University, 0194 Tbilisi, Georgia}

\author{E.~Sanchez\orcidlink{0000-0002-9646-8198}}
\affiliation{CIEMAT, Avenida Complutense 40, E-28040 Madrid, Spain}

\author{D.~Schlegel}
\affiliation{Lawrence Berkeley National Laboratory, 1 Cyclotron Road, Berkeley, CA 94720, USA}

\author{M.~Schubnell}
\affiliation{Department of Physics, University of Michigan, 450 Church Street, Ann Arbor, MI 48109, USA}
\affiliation{University of Michigan, 500 S. State Street, Ann Arbor, MI 48109, USA}

\author{H.~Seo\orcidlink{0000-0002-6588-3508}}
\affiliation{Department of Physics \& Astronomy, Ohio University, 139 University Terrace, Athens, OH 45701, USA}

\author{J.~Silber\orcidlink{0000-0002-3461-0320}}
\affiliation{Lawrence Berkeley National Laboratory, 1 Cyclotron Road, Berkeley, CA 94720, USA}

\author{D.~Sprayberry}
\affiliation{NSF NOIRLab, 950 N. Cherry Ave., Tucson, AZ 85719, USA}

\author{G.~Tarl\'{e}\orcidlink{0000-0003-1704-0781}}
\affiliation{University of Michigan, 500 S. State Street, Ann Arbor, MI 48109, USA}

\author{B.~A.~Weaver}
\affiliation{NSF NOIRLab, 950 N. Cherry Ave., Tucson, AZ 85719, USA}

\author{R.~Zhou\orcidlink{0000-0001-5381-4372}}
\affiliation{Lawrence Berkeley National Laboratory, 1 Cyclotron Road, Berkeley, CA 94720, USA}

\author{H.~Zou\orcidlink{0000-0002-6684-3997}}
\affiliation{National Astronomical Observatories, Chinese Academy of Sciences, A20 Datun Road, Chaoyang District, Beijing, 100101, P.~R.~China}

\collaboration{DESI Collaboration}

%\author[a,b,c,d]{Benedict Bahr-Kalus}
%\author[d,e]{David Parkinson}
%\author[d,e]{Kushal Lodha}
%\author[f]{Eva-Maria Mueller}
%\author[g]{Edmond Chaussidon}
%\author[h]{Arnaud de Mattia}
%\author[i]{Daniel Forero-Sanchez}
%\affiliation[a]{INAF – Istituto Nazionale di Astrofisica, Osservatorio Astrofisico di Torino,\\ 10025 Pino Torinese, Italy}
%\affiliation[b]{INFN – Istituto Nazionale di Fisica Nucleare, Sezione di Torino, 10125 Torino, Italy}
%\affiliation[c]{Dipartimento di Fisica, Universit\`a degli Studi di Torino,\\ Via P.\ Giuria 1, 10125 Torino, Italy}
%\affiliation[d]{Korea Astronomy and Space Science Institute,\\ 776 Daedeok-daero, Yuseong-gu, Daejeon 34055, Republic of Korea.}
%\affiliation[e]{University of Science and Technology, Daejeon 34113, Republic of Korea}
%\affiliation[f]{Department of Physics and Astronomy, University of Sussex, Brighton BN1 9QH, UK}

% E-mail addresses: only for the corresponding author
%\emailAdd{benedict.bahrkalus@inaf.it}

%\abstract{
\begin{abstract}
The peak of the matter power spectrum, known as the turnover (TO) scale, is determined by the horizon size at the time of matter-radiation equality. 
This scale can serve as a standard ruler, independent of other features in the matter power spectrum, such as baryon acoustic oscillations (BAO). 
Here, we present the first detection of the turnover in the galaxy auto-power spectrum, utilising the distribution of quasars (QSO) and luminous red galaxies (LRG) measured by the Dark Energy Spectroscopic Instrument (DESI) during its first year of survey operations 
in a model-independent manner. To avoid confirmation bias, we first analyse the data using data blinding methods designed for the DESI baryon acoustic oscillation, redshift space distortion and scale-dependent bias signals. We measure the angle-averaged dilation distance $D_\mathrm{V}(z = 1.651) = \left(38.1\pm 2.5\right)r_\mathrm{H}$ from the quasars and $D_\mathrm{V}(z = 0.733) = \left(21.8\pm 1.0\right)r_\mathrm{H}$ from the LRG sample in units of the horizon $r_\mathrm{H}$ at the matter-radiation-equality epoch. Combining these two constraints and assuming a flat $\Lambda$CDM model with three standard neutrino species, we can translate this into a constraint of $\Omega_\mathrm{m}h^2 = 0.139^{+0.036}_{-0.046}$. We can break the $\Omega_\mathrm{m}$-$H_\mathrm{0}$ degeneracy with low-redshift distance measurements from type-Ia supernova (SN) data from Pantheon+, we obtain a sound-horizon free estimate of the Hubble-Lema\^itre parameter of $H_0=\HOfromTOandSNeBF$, consistent with sound-horizon dependent DESI measurements. On the other hand, combining the DESI BAO and TO, we find a truly DESI-only measurement of $H_0=\HOfromTOandBAOBF$, in line with DESI-only full-shape results where the sound-horizon scale is marginalised out. This discrepancy in $H_0$ can be reconciled in a $w_0w_a$CDM cosmology, where the combination of DESI BAO and TO data yields $H_0 = 66.5\pm 7.2\;\mathrm{km/s/Mpc}$.
\end{abstract}
\pacs{98.80.Es}
%\begin{document}
\maketitle
\flushbottom

\section{Introduction}
\label{sec:intro}

The distribution of matter in the Universe is observed through the distribution of tracer particles, such as galaxies, quasi-stellar objects (QSOs) or Lyman-alpha absorption systems at low-redshift or hot and cold spots in the cosmic microwave background (CMB) at high-redshift. The formation of structures during two epochs is linked through gravitational collapse, as the seeds of structure formation are fixed during the relativistic era. Matter perturbations have only just begun to grow when the CMB photons are emitted during recombination and the size of fluctuations $\frac{\delta\rho}{\rho}\sim 10^{-5}$.  This early era of linear growth, where cold dark matter (CDM) perturbations can grow without the restoring force of photon pressure while the baryon perturbations experience acoustic oscillations, starts at the epoch of matter-radiation equality.  At this epoch the two fluids have equal energy density, and so any (small-scale) perturbation that entered the horizon before this epoch is retarded in its growth, whereas the larger perturbations that enter after experience continual enhancement. As such, the horizon scale at matter-radiation equality appears in the matter density power spectrum as a peak.

This matter-radiation equality scale (also known as the power spectrum turnover) is fixed in comoving coordinates, similar to the baryon acoustic oscillation (BAO) sound horizon, and can be measured in the same way at low redshift, using galaxy redshift surveys. Since the scale can be measured in a (relatively) cosmologically-independent way, it can be used as a standard ruler to probe the expansion history of the Universe. The scale is calibrated by the ratio of the relativistic vs non-relativistic energy densities, where the total energy density of the relativistic material in the Universe includes the sum of the photon and neutrino energy densities.

Making an accurate detection of the equality scale from the matter power spectrum requires a much larger volume than is needed for the BAO. A first  indication was attempted using data from the WiggleZ Dark Energy Survey \cite{Poole:WiggleZTO}, with an effective survey volume of only $0.84\; (\mathrm{Gpc}/h)^3$. A more solid measurement of a power spectrum inflexion point was achieved using the eBOSS quasar sample \cite{Bahr-Kalus:eBOSSTO}, however, there was no clear evidence for a positive large-scale slope of the power spectrum. The power spectrum turnover has recently been detected in the projected clustering of quasars in the Gaia-unWISE quasar catalogue, Quaia, and its cross-correlation with CMB lensing data from Planck \cite{Alonso:TO}.

In this work, we aim to improve on our eBOSS power spectrum turnover measurement with a QSO sample with a three times larger effective volume and an additional tracer from the Dark Energy Spectroscopic Instrument (DESI), a multi-object fibre-fed spectroscopic system with 5000 fibers covering the focal plane with a $\sim 3^o$ field of view \cite{DESI2016b.Instr,FocalPlane.Silber.2023,Corrector.Miller.2023,FiberSystem.Poppett.2024}. The spectrograph is mounted on the focal plane of the Mayall Telescope at Kitt Peak National Observatory, Arizona \cite{DESI2022.KP1.Instr}. The DESI spectrograph is being used for the large-scale structure galaxy redshift survey that started in 2020, which will measure the spectra of 35 million galaxies over 5 years \cite{DESI2016a.Science}.

The goal of the Dark Energy Spectroscopic Instrument (DESI) is to determine the nature of dark energy through the most precise measurement of the expansion history of the universe ever obtained \cite{Snowmass2013.Levi}. DESI was designed to meet the definition of a Stage IV dark energy survey with only a 5-year observing campaign. Forecasts for DESI \cite{DESI2016a.Science} predict a factor of approximately five to ten improvement on the size of the error ellipse of the dark energy equation of state parameters $w_0$ and $w_a$ relative to previous Stage-III experiments. Here, the constraints on the dark energy parameters are primarily coming from measurements of the BAO standard ruler, which is sensitive to both the angular diameter distance and the Hubble parameter, evaluated at the redshift of the galaxy sample.

The matter-radiation equality scale and the BAO scale differ in the physical mechanisms that generate them, and so are calibrated differently. Recent analyses of DESI BAO \cite{DESI2024.III.KP4,DESI2024.IV.KP6} have either marginalised over this physical sound horizon scale (to use the BAO as relative-size rulers) or else combined the BAO data at low-redshift with measurements of the CMB temperature and polarisation anisotropies at high redshift, which provide very accurate predictions of the sound horizon at the drag epoch \cite{DESI2024.VI.KP7A}. The matter-radiation equality scale can be calibrated without reference to the CMB power spectrum, requiring only a knowledge of the total relativistic energy density of the Universe, which comes from the monopole CMB temperature, plus our understanding of the neutrino sector. Because of this, it can be used to make a measurement of the physical matter density ($\Omega_mh^2$), in contrast to the BAO that are sensitive to the matter density as a fraction of the critical density, $\Omega_m$. Further, our analysis method extracts the matter-radiation equality scale in a model-independent fashion, and (as we will show) our measurements have very small covariance with the DESI BAO measurements.

This work is one of a number of complementary DESI projects that use sound-horizon-independent methods to obtain a measurement of $H_0$. Our companion projects achieve this by either marginalising over the sound horizon 
\cite{DESI:BAO-free}, by using only energy densities \cite{Krolewski:2024jwj}, or by using other power spectrum or correlation function features that depend on the epoch of matter-radiation equality \cite{Brieden:2022heh,Prada:2011uz,DESI:ZeroCrossing}. 

In section \ref{sec:horizonscale} we describe the physics of the matter-radiation scale, and in section \ref{sec:method} we describe the methodology used to measure it. In section \ref{sec:data} we describe the DESI data, as well as the mock catalogues that we used to validate our method and compute the covariances. In section \ref{sec:results} we give the measured values and errors of the matter-radiation scale, and the inferred cosmological parameter constraints. We summarise our findings in section \ref{sec:summary}.

\section{The Horizon Scale at Matter-Radiation Equality}
\label{sec:horizonscale}

The epoch of matter-radiation equality is a major changing point in the history of the Universe due to the change in dynamics, in the growth rate and propagation of the density perturbations. As the expansion decelerates, the Hubble-Lema\^itre rate falls, and primordial perturbations at super-horizon scales can re-enter the horizon, restarting their evolution. However, for those perturbations that already re-enter during radiation domination, the pressure of the relativistic mass-energy prevents gravitational collapse. It is only when non-relativistic matter starts to dominate the dynamics that density perturbations can grow. The specific horizon size at the epoch of matter-radiation equality is a significant scale in the structure formation history of the Universe, depending only on its history up until then, and unaffected by any future change in the dynamics (e.g. curvature, dark energy). It is given by
\begin{equation}
    r_\mathrm{H} = c\int_0^{a_\mathrm{eq}}\frac{\de a}{a^2 H(a)},
    \label{eq:rH}
\end{equation}
where the scale factor at matter-radiation equality $a_\mathrm{eq} = (1 + z_\mathrm{eq})^{-1} = \Omega_\mathrm{r}/\Omega_\mathrm{m}$ can be expressed in terms of the ratio of the radiation and matter density parameters. This simple equation, used in \cite{Bahr-Kalus:eBOSSTO}, becomes slightly more complicated in the presence of massive neutrinos.

\begin{figure}[htbp]
\centering
\includegraphics[width=\columnwidth]{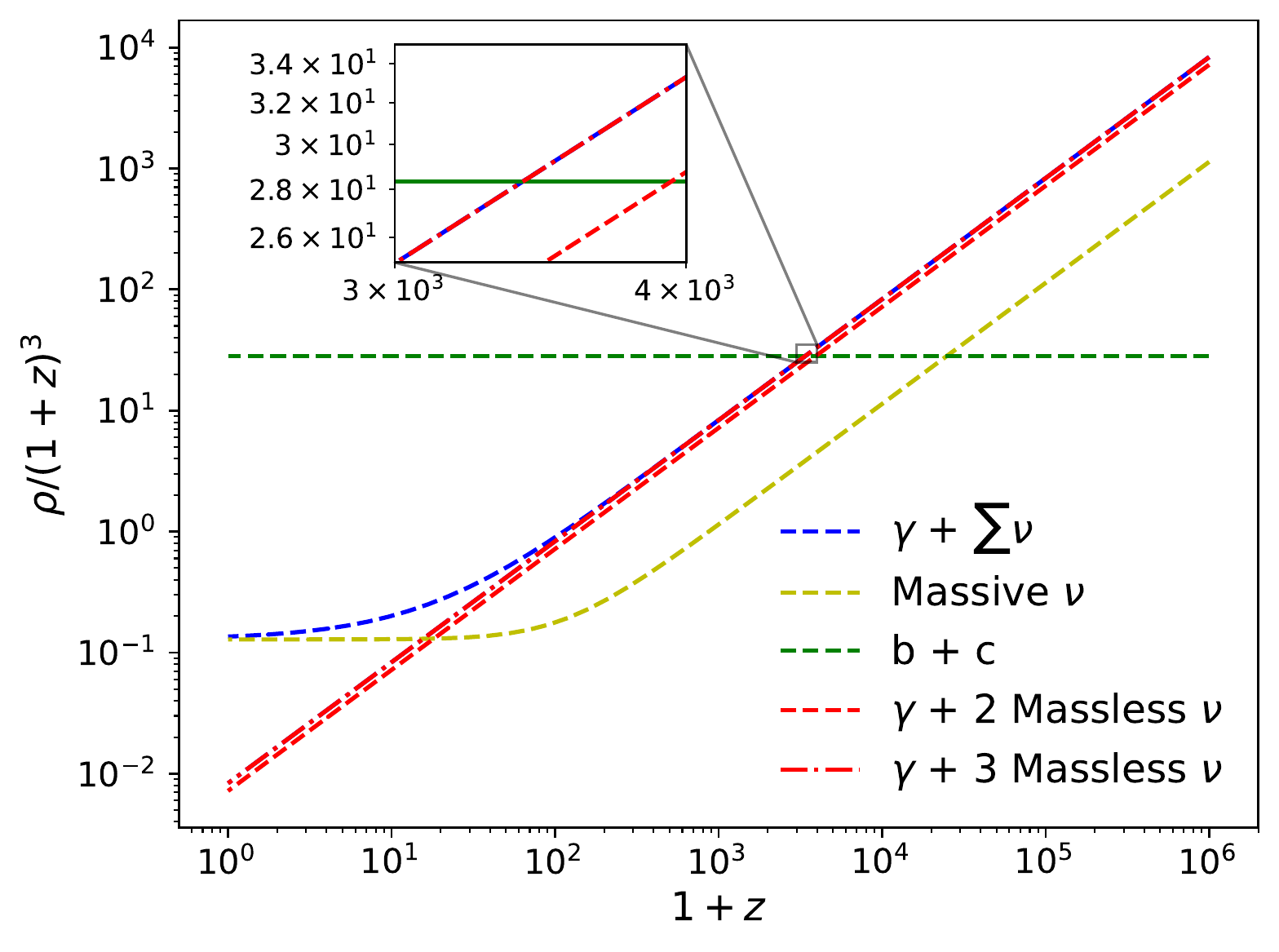}
\caption{The evolution of the mass density of different components as a function of redshift, assuming a single massive neutrino species with mass 0.06 eV (golden line), and two other massless species (combined with the photons in the red dashed line). The equality redshift $z_\mathrm{eq}$ is given by the point where the blue (total relativistic energy density) and green (total non-relativistic energy density) lines coincide. For comparison, we also show the combined neutrino energy density in a cosmology where all neutrinos are massless as a red dot-dashed line. We can observe, that at the matter-radiation-equality epoch, the massive neutrino effectively behaves like a massless neutrino. \label{fig:massdensity-neutrino}}
\end{figure}

For massless neutrinos, the energy density scales as $a^{-4}$ in the same manner as photons. However, massive neutrinos are relativistic in the early Universe but become non-relativistic later, functioning as a form of warm dark matter that slows gravitational collapse. Most cosmological Boltzmann codes treat the massive neutrinos in this manner, as a fluid with some non-constant equation of state. The evolution of the different density values is shown in figure \ref{fig:massdensity-neutrino}.

Since massive neutrinos become non-relativistic at late time, they should be considered as part of the relativistic energy density budget when computing the redshift of matter radiation equality. In practice, this means that when computing this equality redshift, the massive neutrino (or neutrinos) should be treated as though it were massless, and the total effective number of neutrino species $N_{\mathrm{eff}}$ adjusted accordingly. So, the redshift of equality is given by
\begin{equation}
    a_\mathrm{eq} = (1+z_{\mathrm{eq}})^{-1} = \frac{\rho_\gamma + \rho_\nu}{\rho_\mathrm{b}+\rho_{\mathrm{cdm}}} \,,
    \label{eq:aeq}
\end{equation}
where the neutrino density is computed assuming all species are massless,
\begin{equation}
    \rho_\nu = N_{\mathrm{eff}} \frac{7\pi^2}{120}T_\nu^4 \,.
    \label{eq:rho_nu}
\end{equation}
Here the neutrino temperature is
\begin{equation}
   T_\nu = \left(\frac{11}{4}\right)^{-1/3}T_\gamma\,,
   \label{eq:neutrino_temp}
\end{equation}
and the effective number of neutrino species $N_{\mathrm{eff}}$ is equal to the total effective number of neutrino species, including massive and massless.

Considering the early Universe (i.e. $a < a_\mathrm{eq}$) being composed only of photons, effectively relativistic neutrinos, baryonic, and cold dark matter, we can evaluate the integral of \eqref{eq:rH} to express the horizon scale as
\begin{equation}
    r_\mathrm{H} = \frac{2c\left(\sqrt{2} - 1\right)\sqrt{a_\mathrm{eq}}}{H_0\sqrt{\Omega_\mathrm{bc}}}
    \label{eq:rH_func_of_omega}
\end{equation}
which is inversely proportional to the square root of $\omega_\mathrm{bc} = \omega_\mathrm{b} + \omega_\mathrm{cdm}$, where $\omega_i = \Omega_i h^2$ denotes the physical energy density of fluid $i$, and bc denotes baryonic and cold dark matter, explicitly excluding massive neutrinos. 

Following \cite{Eisenstein:EH98}, we can define the equality wavenumber as $k_\mathrm{eq} = a_\mathrm{eq} H(a_\mathrm{eq})$. Under the same assumptions that led to \eqref{eq:rH_func_of_omega}, it is straight-forward to relate this quantity to $r_\mathrm{H}$ as \cite{Prada:2011uz}
\begin{equation}
    k_\mathrm{eq} = 2\left(2 - \sqrt{2}\right)r_\mathrm{H}^{-1}
    \label{eq:keq_vs_rH}
\end{equation}
This $k_\mathrm{eq}$ is approximately but not exactly the turnover scale $k_\mathrm{TO}$. $k_\mathrm{TO}$ has a weak dependence on the baryon density $\omega_\mathrm{b}$. The two scales can be related by the fitting formula \cite{Prada:2011uz} $
    k_\mathrm{TO} = \frac{0.194}{\omega_\mathrm{b}^{0.321}}\left(k_\mathrm{eq}\,\mathrm{Mpc}\,h^{-1}\right)^{0.685 - 0.121\log_{10}\left(\omega_\mathrm{b}\right)}$
to connect the two scales. However, here we are going to take a more accurate approach and find the maximum of fiducial power spectra computed using the \texttt{cosmoprimo} interface\footnote{\url{https://github.com/cosmodesi/cosmoprimo}} to the \texttt{CLASS} Boltzmann solver \cite{Blas:class}.

\section{Methods}
\label{sec:method}
\subsection{Fiducial Cosmology}
\label{sec:fidcosmo}
Like the BAO analyses \cite{DESI2024.III.KP4,DESI2024.IV.KP6}, our analysis here uses a fiducial cosmology to convert redshifts into distances. Thus, our distance measurements are relative to this fiducial cosmology. Our fiducial cosmology matches the average cosmological parameter values from the Planck 2018-$\Lambda$CDM base\_plikHM\_TTTEEE\_lowl\_lowE\_lensing chains \cite{Planck:params}. The key parameters are $\omega_\mathrm{b} = 0.02237$, $\omega_\mathrm{cdm} = 0.1200$, $h = 0.6736$, the effective number of ultra-relativistic species $N_\mathrm{ur} = 2.0328$ and one massive neutrino with $\omega_\mathrm{\nu} = 0.00064420$. This is the fiducial cosmology throughout DESI and also corresponds to the cosmology used when simulating the 25 base AbacusSummit boxes. Computing the power spectrum for this fiducial cosmology using \texttt{cosmoprimo}, we find a peak at
\begin{equation}
    k_\mathrm{TO,fid} = 16.5 h/\mathrm{Gpc}.
\end{equation}
In analogy to the BAO analyses, we define a dilation parameter
\begin{equation}
    \alpha_\mathrm{TO} = \frac{k_\mathrm{TO}}{k_\mathrm{TO,fid}}.
\end{equation}
Even though the turnover scale $k_\mathrm{TO}$ only approximates the equality scale $k_\mathrm{eq}$ (cf. Section \ref{sec:horizonscale}), mock analyses show that
\begin{equation}
    \alpha_\mathrm{TO} = \frac{D_\mathrm{V}(z)}{D_{\mathrm{V,fid}}(z)}\frac{r_\mathrm{H, fid}}{r_\mathrm{H}}
    \label{eq:alphaeq}
\end{equation}
in analogy to the isotropic BAO (where $D_\mathrm{V}(z) = \sqrt[3]{\frac{cz}{H(z)}D_\mathrm{M}^2(z)}$ and $D_\mathrm{M}(z)$ is the comoving distance to redshift $z$) works well to link $k_\mathrm{TO}$ to cosmology, as long as the fiducial cosmology is not unreasonably different from the true cosmology. We show in Appendix \ref{app:alpha_scaling} and figure \ref{fig:alphaTO_scaling} that eq. \eqref{eq:alphaeq} works well for a wide range of parameter values. Whereas the modelling used for our eBOSS analysis \cite{Bahr-Kalus:eBOSSTO} based on a fitting formula to link $k_\mathrm{eq}$ and $k_\mathrm{TO}$ yields biased results in cases where the true cosmology has either a high value of $h$ or a low value of $\omega_\mathrm{cdm}$.

In order to compute the fiducial horizon size $r_\mathrm{H,fid}$, we use \eqref{eq:aeq} - \eqref{eq:neutrino_temp} to first compute the redshift 
\begin{equation}
    z_\mathrm{eq,fid} = 3408
\end{equation}
at which matter-radiation equality happened, which corresponds to a scale factor $a_\mathrm{eq} = 2.93\times 10^{-4}$ and, in turn, to 
\begin{equation}
    r_\mathrm{H,fid} = 112.7\;\mathrm{Mpc}.
\end{equation}

\subsection{Parameterisation}
Close to the turnover, we model the galaxy power spectrum monopole as $\widehat P(k) = K(b_1) P_\mathrm{template}(k) + s_{n, 0}$ where $P_\mathrm{template}$ is parameterised in the following model-independent way:
\begin{equation}
    P_\mathrm{template}(k)=  \begin{cases} P_\mathrm{TO, fid}^{1 - m x^2}  & \text{for } \frac{k}{k_\mathrm{TO,fid}} < \alpha_\mathrm{TO} \\
        P_\mathrm{TO, fid}^{1 - n x^2} &  \text{for } \frac{k}{k_\mathrm{TO,fid}} \geq \alpha_\mathrm{TO},
    \end{cases}
    \label{eq:Poole_Pk}    
\end{equation}
with $P_\mathrm{TO, fid} = P_\mathrm{fid}(k_\mathrm{TO, fid})$, $x = \frac{\ln(k\;\mathrm{Mpc}/h)}{\ln(\alpha_\mathrm{TO}k_\mathrm{TO,fid}\;\mathrm{Mpc}/h)} - 1$ and the Kaiser factor $K(b_1) = b_1^2 + \frac{2}{3}b_1 f_\mathrm{fid} + \frac{1}{5}f_\mathrm{fid}^2$ \cite{Kaiser:RSD} evaluated for the growth rate $f_\mathrm{fid}$ for the fiducial cosmology. We vary in total five parameters in our fits: the scaling parameter $\alpha_\mathrm{TO}$, the slope parameters $m$ and $n$, the linear bias parameter $b_1$, and the residual shot noise $s_{n,0}$.  This parameterisation is equivalent to the one used in measurements of the matter-radiation equality scale using the WiggleZ Dark Energy Survey \cite{Poole:WiggleZTO} and the extended Baryon Oscillation Spectroscopic Survey (eBOSS) Quasar Sample \cite{Bahr-Kalus:eBOSSTO,Bahr-Kalus:eBOSSTOerr}. While \cite{Poole:WiggleZTO,Bahr-Kalus:eBOSSTO,Bahr-Kalus:eBOSSTOerr} also employ two slope parameters $m$ and $n$, here, we express the position of the turnover by a scaling parameter $\alpha_\mathrm{TO}$ and the power spectrum amplitude by the linear bias parameter $b_1$. In this way, we are closer to the BAO implementation in \texttt{desilike} \footnote{\url{https://github.com/cosmodesi/desilike}} and, at the same time, emphasise that \eqref{eq:Poole_Pk} is defined with respect to a fiducial template power spectrum $P_\mathrm{fid}(k)$ which sets the power spectrum peak postion $k_\mathrm{TO,fid}$ and its amplitude $P_\mathrm{TO,fid}$. 

Of course, the parameterisation given in \eqref{eq:Poole_Pk} breaks down the further we are from the turnover scale due to the BAO, non-linear clustering and other small-scale effects. We could prevent these from biasing our results by either an aggressive scale cut or modelling these effects; however, in the former case, we lose valuable broad-band information, and in the latter, we lose model independence. Therefore, we follow \cite{Bahr-Kalus:eBOSSTO,Bahr-Kalus:eBOSSTOerr} and deproject out the modelling systematic as follows: \begin{enumerate}
    \item We compute the mean of the power spectra of 1000 EZmock realisations $\bar P_\mathrm{mock}(k)$ (cf. section \ref{sec:mocks}). Note that this step differs slightly from the eBOSS analysis \cite{Bahr-Kalus:eBOSSTO, Bahr-Kalus:eBOSSTOerr}, where the fiducial power spectrum computed by \texttt{camb} was used instead. This change ensures that the window function, fibre assignment effects, etc.\ are properly included (cf. section \ref{sec:mocks}).
    \item We fit \eqref{eq:Poole_Pk} to the mean mock power spectrum over the same $k$-range as the data, fixing $\alpha_\mathrm{TO} = 1$ and imposing weights inverse to $\sigma^2(k) = \left(k - k_\mathrm{TO, fid}\right)^2$. While this acts like a variance, this is not any data-informed quantity but rather a choice. The goal of this choice is to estimate the modelling error, and as \eqref{eq:Poole_Pk} parameterises the power spectrum well close to the turnover scale, we chose $\sigma$ to reflect the distance between $k$ and the expected turnover scale. We call the best-fitting power spectrum $P_\mathrm{bf}(k)$.
    \item The $k$-dependence of our modelling inaccuracy can then be described by $f(k) = \bar P_\mathrm{mock}(k) - P_\mathrm{bf}(k)$. We free the amplitude of our modelling systematic by introducing a nuisance parameter $\tau$. The true power spectrum then reads $P_\mathrm{true}(k) = \hat P(k) + \tau f(k)$. Assuming independence between $\hat P(k)$ and $f(k)$, the covariance taking our modelling inaccuracy into account is given by $\hat C(k_1, k_2) = C_\mathrm{true}(k_1, k_2) + \tau^2 f(k_1)f(k_2)$. We can analytically marginalise out the nuisance parameter $\tau$ by taking its limit to infinity when inverting $\hat C(k_1, k_2)$, yielding 
    %\begin{widetext}
    \begin{equation}
        \hat C^{-1}(k, q) =  C_\mathrm{true}^{-1}(k, q) - \Delta C(k,q),
    \end{equation}    
    with
    \begin{equation}
        \Delta C(k,q) = \frac{\sum_{\mathhangeulke, \mathhangeulkyu}C_\mathrm{true}^{-1}(k, \mathhangeulke)f(\mathhangeulke)f(\mathhangeulkyu)C_\mathrm{true}^{-1}(\mathhangeulkyu, q)}{\sum_{\mathhangeulke, \mathhangeulkyu}f(\mathhangeulke)C_\mathrm{true}^{-1}(\mathhangeulke, \mathhangeulkyu)f(\mathhangeulkyu)}.
        \label{eq:moddeproj}
    \end{equation}
    %\end{widetext}
\end{enumerate}

\subsection{Likelihood and Fitting}

In our eBOSS analysis \cite{Bahr-Kalus:eBOSSTO}, we identified a discrepancy in determining $k_\mathrm{TO}$ when adhering to the conventional assumption of a Gaussian likelihood for the power spectrum. To rectify this bias, we opted for an alternative approach by employing a likelihood derived from a Box-Cox transformation \cite{Box:1964} applied to the gamma distribution approximating the hypo-exponential distribution of a binned, window-convolved power spectrum of a Gaussian random field \cite{Wang:distro}. It is important to note that in this approximation, the number of available $k$-modes corresponds to the diagonal elements of the precision matrix $\hat C^{-1}(k, k)$. However, the application of \eqref{eq:moddeproj} diminishes the presumed count of modes, which inadvertently incorrectly enhances the non-Gaussian characteristics of the likelihood at smaller scales. Consequently, we have decided not to use the likelihood model employed in the eBOSS TO analysis and utilise instead a conventional Gaussian likelihood defined as:
\begin{equation}
\chi^2 = \sum_{(k_1, k_2) = k_\mathrm{min}}^{k_\mathrm{max}} \Delta P(k_1)\hat C^{-1}(k_1, k_2)\Delta P(k_2),
\label{eq:chi2Gauss}
\end{equation}
where $\Delta P(k) = \tilde P(k) - P_\mathrm{conv}(k)$. We will demonstrate in Section \ref{sec:mocks} that for DESI, this approach provides an unbiased measurement of the turnover scale.

We choose of $k_\mathrm{min} = 0.004\;h/\mathrm{Mpc}$ and $k_\mathrm{max} = 0.2\; h/\mathrm{Mpc}$ to be the same values as in the eBOSS turnover analysis \cite{Bahr-Kalus:eBOSSTO}. Our choice of $k_\mathrm{min}$ is slightly more conservative than in the DESI primordial non-Gaussianity (PNG) analysis \cite{ChaussidonY1fnl} which uses the same data. Larger scales are impacted by geometrical effects and, more importantly, imperfect correction of imaging systematics. On the other hand, our $k_\mathrm{max}$ is significantly larger than $k_\mathrm{max} = 0.08\,h/\mathrm{Mpc}$ chosen for the PNG measurement \cite{ChaussidonY1fnl}. We can use this larger value because our method down-weights these small scales, effectively using only the broadband information from the power spectrum.

We show in Figure \ref{fig:kmax_vary} the posterior contours of turnover parameters obtained from minimising \eqref{eq:chi2Gauss} for the mean power spectrum of DESI Y1 LRG and QSO EZmock realistations (cf. Section \ref{sec:mocks}), assuming the PNG analysis $k_\mathrm{max} = 0.08\,h/\mathrm{Mpc}$, the eBOSS turnover  analysis $k_\mathrm{max} = 0.2\,h/\mathrm{Mpc}$, and an intermediate $k_\mathrm{max}=0.15\,h/\mathrm{Mpc}$. As expected, we see significant differences in the marginalised 1D posterior of the small-scale slope parameter $n$. However, the quantities we care about -- the fraction of the posterior volume with $m>0$ and the value of $\alpha_\mathrm{TO}$ -- remain remarkably stable. The probability of not detecting the turnover feature (i.e. $\mathcal{P}(m\leq 0)$) is remarkably stable when changing $k_\mathrm{max}$, diminishing from 21.6\% in mock LRGs at $k_\mathrm{max}=0.08\,h/\mathrm{Mpc}$ to 21.1\% at higher $k_\mathrm{max}$, and changing from 0.06\% at $k_\mathrm{max}=0.08\,h/\mathrm{Mpc}$, to 0.04\% at $k_\mathrm{max}=0.15\,h/\mathrm{Mpc}$ and 0.08\% at $k_\mathrm{max}=0.2\,h/\mathrm{Mpc}$ with mock QSOs. The maximum a posteriori value of $\alpha_\mathrm{TO}$ is also stable under variations of $k_\mathrm{max}$ (LRG: 0.99, 1.002, 0.995; QSO: 1.01, 1.007, 1.004, for $k_\mathrm{max} = [0.08, 0.15, 0.2]\, h/\mathrm{Mpc}$, respectively). What we do notice for $\alpha_\mathrm{TO}$ is that by carving away posterior space using small-scale broadband information on $n$, we reduce the standard deviation on $\alpha_\mathrm{TO}$ from 0.22 to 0.091 and 0.071 with the mock LRGs and from 0.11 to 0.085 and 0.080 with the mock QSOs, again for $k_\mathrm{max} = [0.08, 0.15, 0.2]\, h/\mathrm{Mpc}$. Therefore, we have chosen $k_\mathrm{max} = 0.2\,h/\mathrm{Mpc}$ for our analysis, as it provides tighter yet stable constraints on the parameters of interest.

\begin{figure}[htbp]
\centering
\includegraphics[width=\columnwidth]{kmax_vary_combined_triangle.pdf}
\caption{Constraints on turnover parameters from the mean of DESI Y1 LRG and QSO EZmock power spectra obtained with different values of $k_\mathrm{max}$. The filled contours correspond to the fiducial $k_\mathrm{max} = 0.2\, h/\mathrm{Mpc}$ used throughout this article.\label{fig:kmax_vary}}
\end{figure}

All inferences are performed using the \texttt{emcee} sampler\footnote{\url{https://github.com/dfm/emcee}}\cite{ForemanMackey:emcee} within the \texttt{desilike} framework. A practical outline of our analysis pipeline is available as a Jupyter notebook on the \texttt{cosmodesi} GitHub page.\footnote{\url{https://github.com/cosmodesi/desilike/blob/main/nb/turnover\_examples.ipynb}}

\section{Data}
\label{sec:data}
We make use of data from the first year of spectroscopic observations carried out with the DESI instrument \cite{DESI2016b.Instr}
mounted on the Nicholas U. Mayall Telescope at Kitt Peak National Observatory on  Iolkam Du'ag in Arizona from the 14th of May, 2021, until the 14th of June, 2022.
Each observation field is covered by a ``tile", consisting in a set of targets \cite{DESI:targetting} selected from the photometric catalogues of the 9th public data
release of the DESI Legacy Imaging Surveys \cite{DESI:legacysurveyDR9}\footnote{\url{https://www.legacysurvey.org/dr9/}} and assigned to each of the 5000 fibres in the telescope's focal plane. The observed data are processed by the DESI spectroscopic pipeline \cite{DESI:spectropipeline}. 
The covered tile surface area is about $7,500\;\mathrm{deg^2}$, i.e. roughly half of DESI's expected final coverage of $14,200\;\mathrm{deg^2}$. 
The extragalactic targets are defined to fall into the four following classes: the bright galaxy sample (BGS,
\cite{BGS.TS.Hahn.2023}), luminous red galaxies (LRG, \cite{DESI:LRGs}), emission line galaxies (ELG \cite{DESI:ELGs}), and quasars (QSO \cite{DESI:QSOs}). As we need as much volume as possible to observe the large-scale feature of the power spectrum turnover, we do not consider the BGS sample, extending only to a maximum redshift of $z_\mathrm{max} = 0.6$, for this analysis. Although the ELG sample covers the redshift range of $1.1 < z < 1.6$ (overlapping with both LRGs and QSOs), significant efforts to control systematic effects at scales much larger than the BAO have not yet been fully successful \cite{RosadoMarin:DESIELGsyst}. Hence, we only use the LRG and QSO samples for this study, which we describe in the following subsections.
For a more general and technical overview of the catalogue building, we refer to \cite{DESI:catalogues}.

\begin{figure}[htbp]
\centering
\includegraphics[width=\columnwidth]{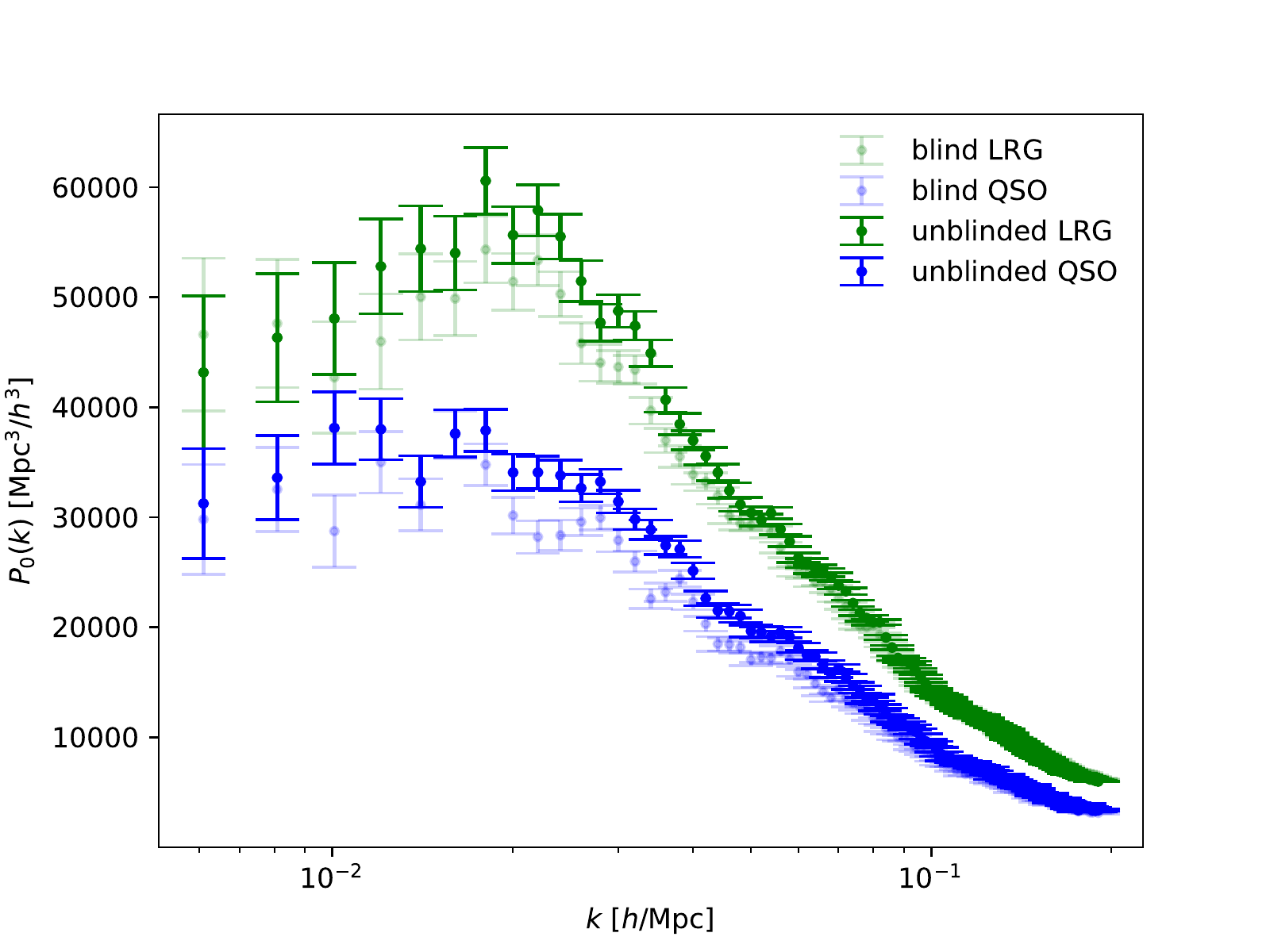}
\caption{DESI Y1 LRG and QSO power spectra. Pale data points and errorbars show the blinded data, whereas opaque ones the data after unblinding. The solid curves represent the best fit (cf. Section \ref{sec:results}). The apparent mismatch between data and best fit at small scales are the effect of the mode deprojection of the modelling uncertainty at these scales far from the power spectrum peak (cf. equation \ref{eq:moddeproj}). }
\label{fig:pk}
\end{figure}

In particular, we use the power spectra and covariance matrices (cf. Figure \ref{fig:pk}) generated for constraints on the scale-dependent bias signature of local-type PNG \cite{ChaussidonY1fnl}. We will briefly summarise how they are obtained but refer the reader to the primordial non-Gaussianity paper for a detailed description of how they have been obtained. Unlike the DESI key analyses \cite{DESI2024.III.KP4,DESI2024.IV.KP6,DESI2024.VI.KP7A,DESI2024.VII.KP7B}, we make use of the full available redshift ranges of the LRG and QSO samples, i.e. we do not subdivide the LRG sample into redshift bins, and we include QSOs at redshifts higher than $z = 2.1$ which are principally targetted for Ly-$\alpha$ forest analyses. 

The QSO sample \cite{DESI:QSOs} used here thus contains 1,189,129 objects (thus about 40 per cent larger than the standard QSO clustering sample with 856,652 objects and more than three times as large as the eBOSS QSO sample \cite{Ross:eBOSScatalogue} with 343,708 objects used in \cite{Bahr-Kalus:eBOSSTO,Bahr-Kalus:eBOSSTOerr}) with redshift values ranging from 0.8 to 3.1. The effective redshift is $z_\mathrm{eff} = 1.651$ \cite{ChaussidonY1fnl} when using Feldman-Kaiser-Peacock (FKP) weights \cite{Feldman:FKPweights}. Note that this is lower than when using the optimal redshift weights applied in the DESI PNG analysis \cite{ChaussidonY1fnl}. 

Apart from containing more objects than the eBOSS QSO sample, the DESI QSO sample covers a larger area, as well as a higher density due to the high priority given to QSOs in the fibre assignment process. This results in an effective volume at the turnover scale about three times larger than that of the eBOSS QSO sample. 

The LRG sample used here combines all three LRG subsamples used in the DESI key analyses \cite{DESI2024.III.KP4,DESI2024.IV.KP6,DESI2024.VI.KP7A,DESI2024.VII.KP7B}. Its effective redshift amounts to $z_\mathrm{eff} = 0.733$ using standard FKP weights \cite{ChaussidonY1fnl}.

Variations in completeness, imaging systematics, and spectroscopic efficiency are accounted for using additional weights as described in \cite{Ross:DESICat}. As imaging systematics are crucial to this analysis, the imaging systematics weights are further improved using the regression code \texttt{regressis} \cite{Chaussidon:regressis_and_DESIQSOsel,ChaussidonY1fnl}.\footnote{\url{https://github.com/echaussidon/regressis}} However, as shown in figure 25 of \cite{ChaussidonY1fnl}, the systematic weights primarily affect super-equality-horizon scales for the DESI QSO and LRG samples. This indicates that the $\alpha_\mathrm{TO}$ measurement is robust against imaging systematics. However, imaging systematics can flatten the power spectrum at these scales or even cause an upturn, which may result in an underestimation of the slope parameter $m$ and, in turn, we might be prone to underestimating our detection probability.

We avoid confirmation bias by first testing and performing our analysis on blinded data. The blinding strategy and its validation is described in detail in \cite{Chaussidon:blinding,Andrade:blinding}. DESI is blinded at the catalogue level. At ultra-large scales, the dominant blinding contribution comes from imposing a randomly chosen value of $f^\mathrm{blind}_\mathrm{NL} \in \left[-15, 15\right]$ onto a set of weights such that one will measure $f_\mathrm{NL}^\mathrm{obs} = f_\mathrm{NL}^\mathrm{cat} + f^\mathrm{blind}_\mathrm{NL}$, where $f_\mathrm{NL}^\mathrm{cat}$ is the local PNG parameter measured from the unblinded catalogue. To prevent unblinding without knowing $f^\mathrm{blind}_\mathrm{NL}$, the completeness weights available to the collaboration before unblinding were multiplied by the blinding weights. The blinded power spectrum is shown in pale along their unblinded equivalents in Figure \ref{fig:pk}. We show in section \ref{sec:mocks_blinding} using mocks that the $f_\mathrm{NL}$ blinding strategy is also efficient for this turnover analysis.

Being commensal with the DESI PNG analysis \cite{ChaussidonY1fnl}, our power spectra have been obtained using the Yamamoto estimator \cite{Yamamoto:estimator} implemented in \texttt{pypower}\footnote{\url{https://github.com/cosmodesi/pypower}, based on \cite{Hand:pypower}} even though we only consider the power spectrum monopole here. FKP weights \cite{Feldman:FKPweights}
\begin{equation}
    w_\mathrm{FKP}(\mathbf{x}) = \frac{1}{1 + \bar n(\mathbf{x})P_0}
\end{equation}
are applied to minimise the uncertainty of the power spectrum measurement, where $P_0$ is the power spectrum at the scale of interest. With the equality scale as our scale of interest, $P_0 = 3\times 10^4\;\mathrm{Mpc}^3/h^3$ and $P_0 = 5\times 10^4\;\mathrm{Mpc}^3/h^3$ are chosen for the QSO and LRG power spectrum measurement, respectively. \texttt{pypower} automatically subtracts Poissonian shot noise. Thus, our nuisance parameter $s_{n,0}$ presents the residual shot noise.

\subsection{Mock Realisations}
\label{sec:mocks}
We make use of two sets of simulations: \textit{EZmocks} \cite{Chuang:EZmocks} and \textit{Abacus} \cite{Maksimova:AbacusSummit,Garrison:Abacus}. EZmocks use Gaussian random fields to initialise density perturbations in the early Universe. The evolution of these density perturbations is modelled under the assumption of the Zel'dovich approximation and an effective bias model, which is computationally less intensive than full N-body simulations. This methodology balances computational efficiency and accuracy, particularly on these ultra-large scales that can be well described by linear theory, allowing the creation of 1000 realisations per galactic hemisphere with a box size of $\left(6\; \mathrm{Gpc}/h\right)^3$. 
We use two sets of realisations of EZmocks. For most applications, we use the mocks described in detail in section 3.3.2 of \cite{ChaussidonY1fnl}. When estimating correlations with other DESI measurements, in particular the BAO, we use EZmock realisations described in section 3.2 of \cite{DESI2024.III.KP4}.

Abacus mocks are constructed on the foundation of the AbacusSummit simulations \cite{Maksimova:AbacusSummit,Garrison:Abacus}, a state-of-the-art set of cosmological N-body simulations that use an advanced GPU-accelerated codebase. These simulations are designed to produce high-resolution outputs for large cosmological volumes, making them particularly suited for DESI's vast survey footprint and stringent statistical requirements. With up to trillions of particles, AbacusSummit enables precise modelling of the dark matter density field and the resulting halo distribution. The simulations cover volumes comparable to or larger than the DESI survey, ensuring that cosmic variance is minimised and rare structures are well-sampled. Abacus mocks incorporate advanced halo modelling techniques, such as the Halo Occupation Distribution (HOD), to populate halos with galaxies consistent with DESI's target selection. Due to their higher computational cost, we only have 25 Abacus mocks each with a box size of $\left(2\; \mathrm{Gpc}/h\right)^3$ available. 

\subsection{The Covariance Matrix}

To obtain the covariance matrix of the power spectrum, we repeat all steps initially taken to compute the data power spectra and apply them to all 1000 EZmock realisations that have also been used in the covariance estimation for the DESI PNG analysis \cite{ChaussidonY1fnl}. This process yields the power spectrum $P_s(k)$ for each simulation $s$, with the average denoted as $\bar P(k)$. We compute the sample covariance with the formula
\begin{equation}
    C_\mathrm{samp}(k, q) = \frac{\sum_s\left[P_s(k) - \bar P(k)\right]\left[P_s(q) - \bar P(q)\right]}{n_\mathrm{m} - 1} 
\end{equation}
where $n_\mathrm{m} = 1000$ represents the number of mock realisations. This provides an unbiased estimate of the true covariance matrix $C_\mathrm{true}(k_1, k_2)$. Given that the inverse covariance matrix follows an inverse Wishart distribution, $C_\mathrm{samp}^{-1}(k_1, k_2)$ is only a biased estimate of $C_\mathrm{true}^{-1}(k_1, k_2)$. To obtain an unbiased estimate, we multiply $C_\mathrm{samp}^{-1}(k_1, k_2)$ by the Hartlap factor \cite{Hartlap:factor}:
\begin{equation}
    C_\mathrm{H}^{-1}(k_1, k_2) = \frac{n_\mathrm{m} - n_\mathrm{d} - 2}{n_\mathrm{m} - 1}C_\mathrm{samp}^{-1}(k_1, k_2),
\end{equation}
where $n_\mathrm{d}$ is the number of data points. Furthermore, when using $C_\mathrm{H}^{-1}(k_1, k_2)$ instead of the unknown $C_\mathrm{true}(k_1, k_2)$ in any parameter inference, the likelihood becomes wider due to the marginalisation over possible values of $C_\mathrm{true}(k_1, k_2)$. This effect can approximately be accounted for by the Percival factor \cite{Percival:covariancefactor}:
\begin{widetext}
\begin{equation}
    C_\mathrm{PH}^{-1}(k_1, k_2) = \frac{2 + (n_\mathrm{m} - n_\mathrm{d} - 1)(n_\mathrm{m} - n_\mathrm{d} - 4) + (n_\mathrm{m} - n_\mathrm{d} - 2)(n_\mathrm{p} + 1)}{(n_\mathrm{m} - n_\mathrm{d} - 1)(n_\mathrm{m} - n_\mathrm{d} - 4) + (n_\mathrm{m} - n_\mathrm{d} - 2)(n_\mathrm{d} - n_\mathrm{p})}C_\mathrm{H}^{-1}(k_1, k_2),
\end{equation}
\end{widetext}
where $n_\mathrm{p} = 5$ is the number of parameters. Finally, we substitute $C_\mathrm{true}$ with $C_\mathrm{PH}$ in eq. \eqref{eq:moddeproj} to obtain the covariance matrix used to measure the turnover. 

Note that in the DESI full shape analysis \cite{DESI2024.VII.KP7B}, the covariance matrix from the EZmocks is rescaled to match the analytical prediction from RascalC \cite{KP4s7-Rashkovetskyi}. This is necessary as the EZmocks used in \cite{DESI2024.VII.KP7B} emulate the fibre assignment of DESI, which in turn underestimates the covariance. Here, we use the EZmocks of the DESI PNG measurement \cite{ChaussidonY1fnl}, which ignore the impact of the fibre assignment. While this introduces inaccuracies at small scales that are not relevant for us, it has been shown that it does not underestimate the covariance \cite{ChaussidonY1fnl}.

\subsection{The Window Matrix and the Radial Integral Constraints}
\label{sec:window}
In power spectrum measurements, a window matrix formalism usually accounts for the survey's geometry, masked regions, and incompleteness, ensuring accurate modelling of the power spectrum \cite{Beutler:window}. The survey selection function, $W(x)$, describes the observed fraction of the density field. The convolved power spectrum is computed as a matrix multiplication between the unconvolved power spectrum multipoles and the window matrix, as expressed in Equation (4.8) of the DESI PNG paper \cite{ChaussidonY1fnl}.

We make again use of the window matrix of ref. \cite{ChaussidonY1fnl}. For the monopole ($\ell = 0 $), the window matrix is derived using the random catalogue and normalised appropriately to avoid bias. The implementation utilises \texttt{pypower} for computing the window matrix and \texttt{desilike} for the convolution with the theoretical model. Wide-angle corrections are included in the window matrix at first order.

The redshift distribution used to generate random catalogues is typically inferred directly from the data catalogue using the shuffling method. This approach nulls radial modes in the measured power spectrum, leading to the Radial Integral Constraint (RIC, \cite{deMattia:radintconst}), which introduces anisotropic and scale-dependent effects. The DESI PNG paper \cite{ChaussidonY1fnl} quantifies this contribution through an additive correction to the power spectrum or a multiplicative modification to the window function, $\mathcal{W}\rightarrow \mathcal{W}-\mathcal{W}^\mathrm{RIC}$, ensuring that the RIC is properly accounted for without biasing the measurement of $f_\mathrm{NL}^\mathrm{loc}$.

Ref. \cite{ChaussidonY1fnl} demonstrated that the RIC correction affects both the monopole and quadrupole, with a suppression of power on large scales that can bias $f_\mathrm{NL}^\mathrm{loc}$ measurements if not corrected. For the monopole, this effect can be modelled by decreasing the effective value of $f_\mathrm{NL}^\mathrm{loc}$. The validity of this correction for $f_\mathrm{NL}^\mathrm{loc}$ has been confirmed using 100 pairs of mock catalogues with shuffled and unshuffled randoms and through a blinded procedure, validating its robustness across different power spectrum shapes. Using the same set of shuffled and unshuffled mock realisations as ref. \cite{ChaussidonY1fnl}, we tested that the position of TO is unaffected by the RIC. However, the RIC affects the slope $m$ of super-equality-horizon scales and the uncertainty of $\alpha_\mathrm{TO}$. Hence, we adopt the RIC corrected window matrix from the DESI PNG analysis \cite{ChaussidonY1fnl}.

Finally, since the covariance matrix obtained from shuffled mocks is consistent with that derived from unshuffled mocks, we can estimate the covariance from the full set of 1000 unshuffled mocks. These steps ensure that both the window matrix and RIC effects are correctly incorporated, allowing for unbiased parameter inference in our analysis.

\subsection{Validation of Blinding}
\label{sec:mocks_blinding}
The DESI blinding strategy \cite{Chaussidon:blinding,Andrade:blinding} was not designed with a turnover measurement in mind. However, the turnover scale lies between the scales relevant for the PNG analysis and the BAO scale, so, here, we test whether the blinding developed for these measurements also blinds the turnover measurement. Values of $f_\mathrm{NL}^\mathrm{loc}$ that are consistent with Planck 18 data \cite{Planck:png} only cause sub-per cent shifts in the turnover scale \cite{Cunnington:HITO}. However, the DESI blinding allows for three times larger values, making the bias already scale-dependent at the turnover scale. 

\begin{figure}[htbp]
    \centering
    \includegraphics[width=\columnwidth]{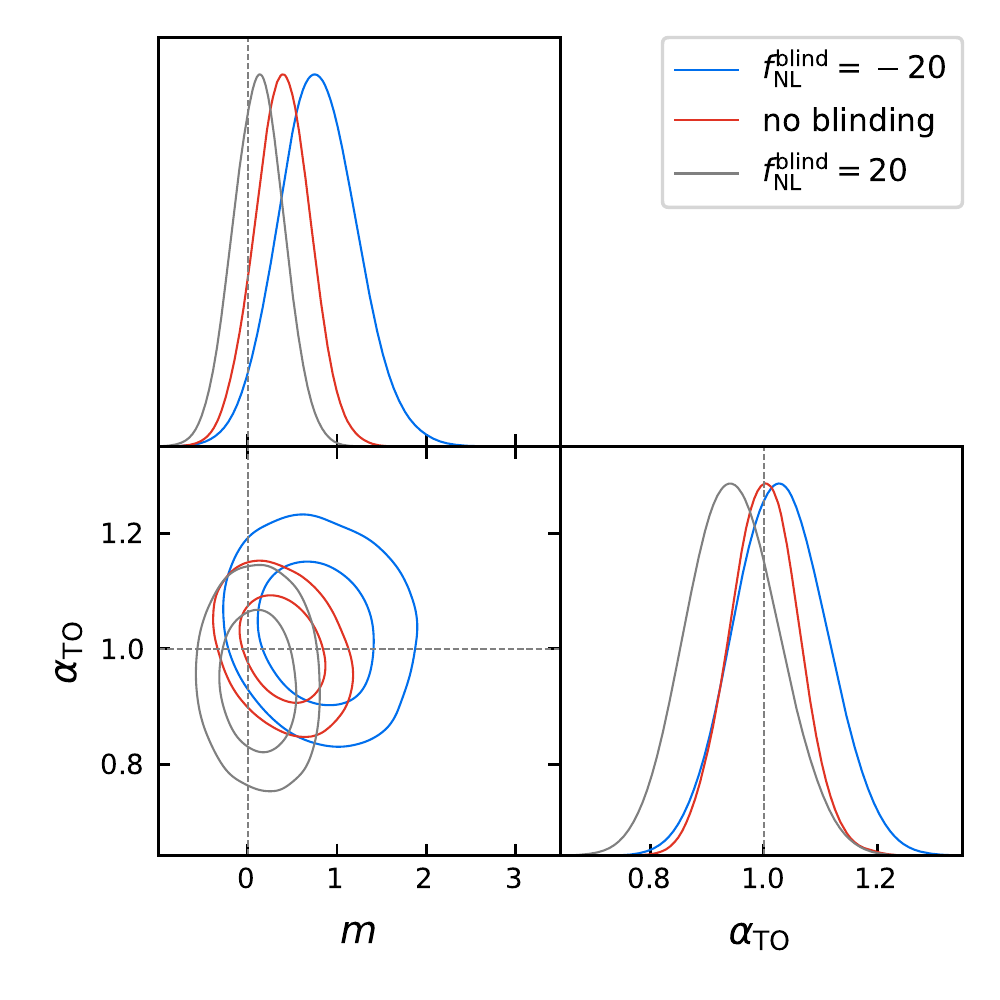}
    \qquad
    \includegraphics[width=\columnwidth]{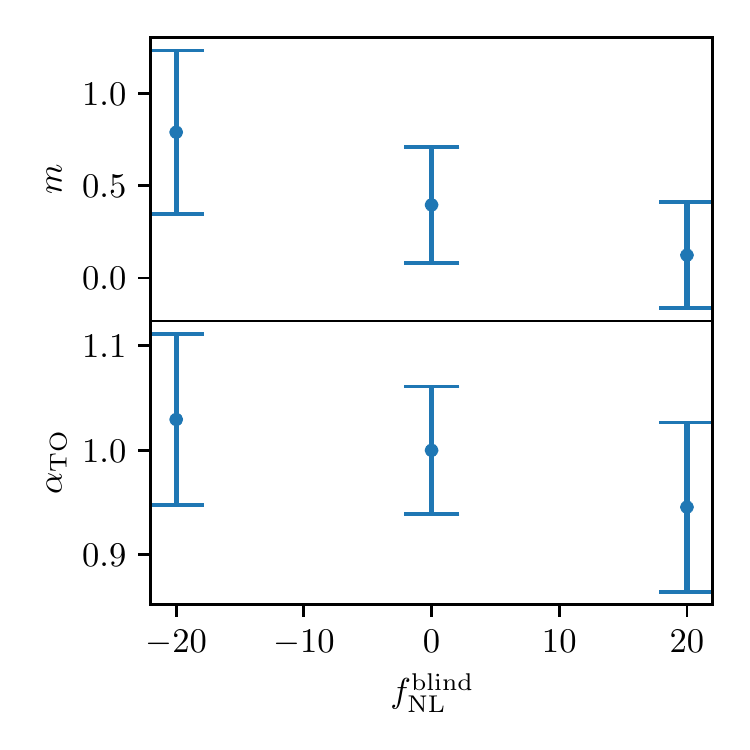}    
    \caption{\textit{Top:} $\alpha_\mathrm{TO}$-$m$ contours obtained from the same Abacus LRG mock without any blinding applied (red), and with a blinding corresponding to $f_\mathrm{NL}^\mathrm{blind} = \pm 20$ applied. \textit{Bottom:} The dependence of the best-fitting values of $m$ and $\alpha_\mathrm{TO}$ and their uncertainties as a function of $f_\mathrm{NL}^\mathrm{blind}$.}
    \label{fig:blinding_dependence}
\end{figure}

Using a set of mocks generated from a single Abacus realisation but with different values of $f_\mathrm{NL}^\mathrm{blind}, w_0^\mathrm{blind}$ and $w_a^\mathrm{blind}$, we measure the $m$, $n$, $\alpha_\mathrm{TO}$, $b_1$, and $s_{n,0}$ posterior distributions of each mock, as shown in figure \ref{fig:blinding_dependence} for variation in $f_\mathrm{NL}^\mathrm{blind}$. We do not see any significant change in the best-fitting parameters as $w_0^\mathrm{blind}$ and $w_a^\mathrm{blind}$ are changed. Nonetheless, the PNG blinding works well also for the turnover measurement presented in this work.

\subsection{Correlation between Quasar and Luminous Red Galaxies Samples}
\label{sec:int_crosscorr}

\begin{figure}[htbp]
    \centering
    \includegraphics[width=\columnwidth]{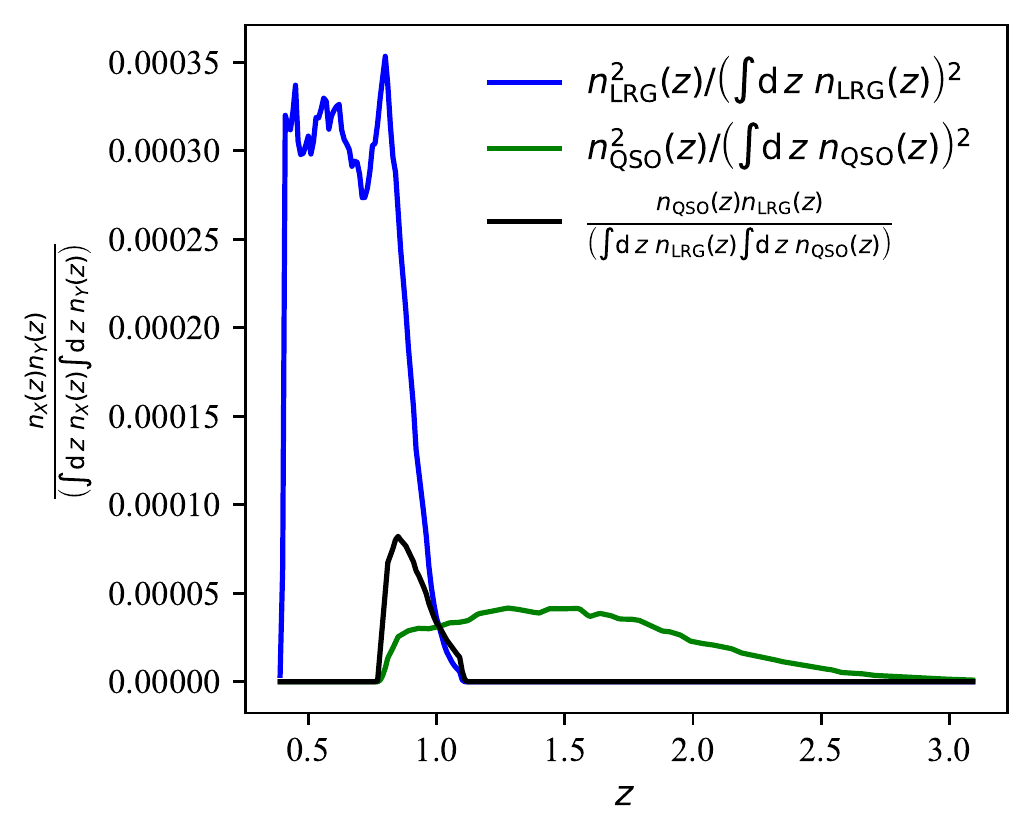}
    \caption{Products of the normalised redshift distributions of the LRG and QSO samples. \label{fig:nn_LRG_vs_QSO}}
\end{figure}

The LRG and QSO samples overlap in the redshift range of $0.8 < z < 1.1$. To perform joint inference using the turnover measurement from both samples, we need to estimate the correlation between the two samples. We can model the cross-power spectrum between two samples as
\begin{equation}
    P_{ij}(k) = \frac{\int \operatorname{d}z\  p_{i}(z)p_{j}(z) P_\mathrm{m}(k,z)}{\int \operatorname{d}z\  n_{i}(z)\int \operatorname{d}z\  n_{j}(z)},
\end{equation}
with $p_i(z) = n_i(z) K_i(z)$, and $n_i(z)$ and $K_i(z)$ denoting the number density and Kaiser factor (cf. below equation \ref{eq:Poole_Pk}) of tracer $i$ at redshift $z$, as well as $P_m(k)$ being the underlying matter power spectrum which is common to both tracers. A rough estimate for the correlation between the two samples can be obtained by
%\begin{widetext}
\begin{align}
    \rho &= \frac{P_\mathrm{LRG,QSO}(k)}{\sqrt{P_\mathrm{LRG}(k)P_\mathrm{QSO}(k)}} \nonumber \\
    &= \frac{\int \operatorname{d}z\  p_\mathrm{LRG}(z)p_\mathrm{QSO}(z) D^2(z)}{\sqrt{\int \operatorname{d}z\  p_\mathrm{LRG}^2(z) D^2(z)}\sqrt{\int \operatorname{d}z\  p_\mathrm{QSO}^2(z) D^2(z)}},
    \label{eq:cross_corr}
\end{align}
%\end{widetext}
where we used $P_\mathrm{m}(k,z) = D^2(z) P_\mathrm{m}(k, 0)$ in the second equality. Adopting the redshift distributions from \cite{ChaussidonY1fnl}, illustrated in figure \ref{fig:nn_LRG_vs_QSO}, we evaluate equation \eqref{eq:cross_corr} and find a percent-level correlation between the LRG and QSO clustering. The full impact on the measurement of $\alpha_\mathrm{TO}$ would be best estimated in mock data. However, the EZmocks built for the sample selections used in \cite{ChaussidonY1fnl} and this work are obtained from different realisations of the underlying matter density field. We, thus, make the conservative choice of estimating the correlation between the LRG and QSO $\alpha_\mathrm{TO}$ measurements from the EZmock realisations used in the DESI direct tracer BAO analysis \cite{DESI2024.III.KP4}. This is conservative because the QSO BAO mocks only extend up to redshift $z = 2.1$ and not $z = 3.1$ as in this analysis. When computing the sample covariance, we find a small correlation coefficient of $0.062$, which we include in the cosmological parameter estimation presented in section \ref{sec:TO_as_std_ruler}. The correlation matrix, along with BAO parameter correlations, is presented in figure \ref{fig:to_bao_corr}.

\section{Results}
\label{sec:results}
\subsection{Measurement of the Turnover Scale}
\begin{figure}[htbp]
\centering
\includegraphics[width=\columnwidth]{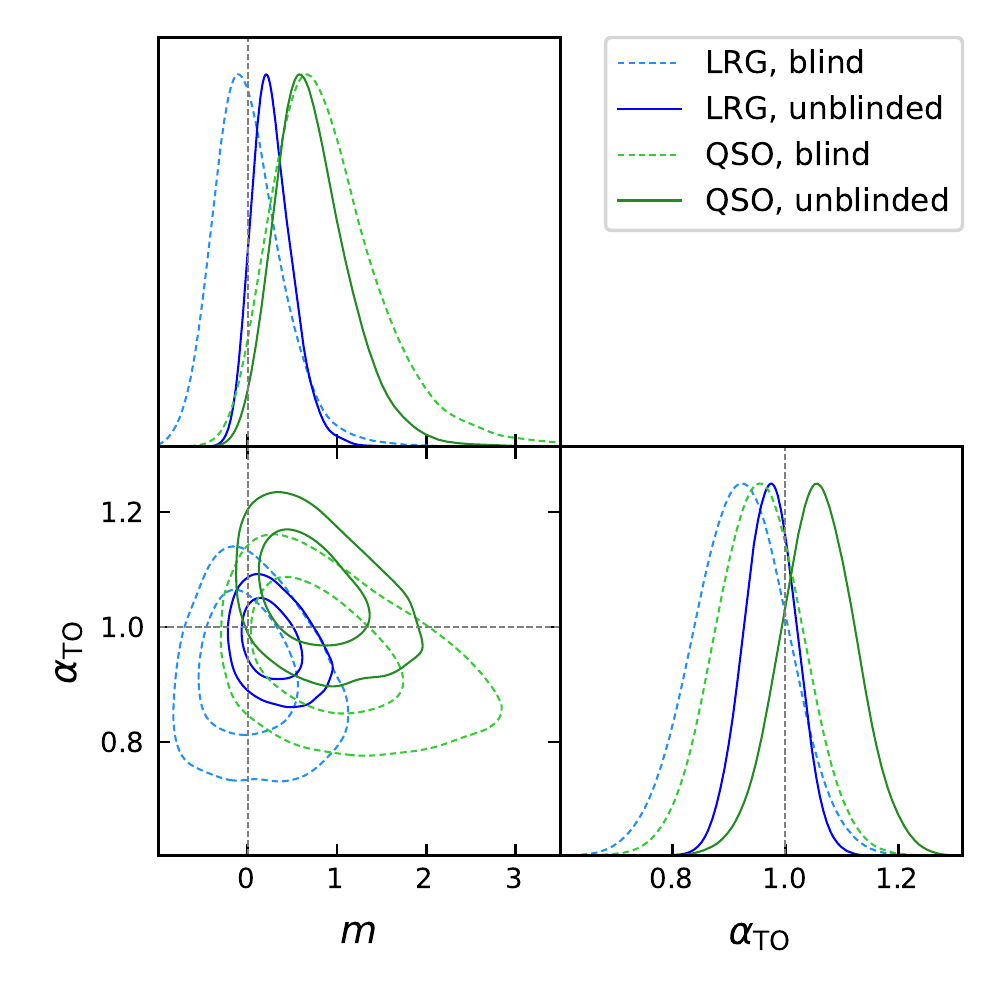}
\caption{$\alpha_\mathrm{TO}$ - $m$ contours from the blind (dashed) and unblinded (solid) LRG and QSO catalogues.\label{fig:combined_contours}}
\end{figure}

We present the model-independent results of our turnover measurement in table \ref{tab:mcmc_output}. The power spectrum can be parameterised well around the turnover by the model-independent parameterisation of equation \eqref{eq:Poole_Pk} as we find $\chi^2_\mathrm{min} = 81.2$ and $\chi^2_\mathrm{min} = 112.8$ from the QSOs and LRGs, respectively. Having used 93 $k$-bins in both fits and with 5 free parameters, this corresponds to a reduced $\chi^2$ of 0.92 from the QSOs and 1.28 from the LRGs. After unblinding, we detect the turnover with a 90 per cent probability in the LRG power spectrum and a 98 per cent probability in QSO data. These probabilities are estimated as the posterior volume with $\mathcal{P}(m>0)$. We show the $\alpha_\mathrm{TO}$-$m$ posterior contours and their respective marginalised 1D distributions in figure \ref{fig:combined_contours}. As the $m$ posterior distribution is non-Gaussian with a pronounced tail towards large positive values, we also list the $m$ posterior means in table \ref{tab:mcmc_output} alongside its best-fitting value. We notice that the contours have shrunk after unblinding, a behaviour we also observe in the mocks (cf. figure \ref{fig:blinding_dependence}). In both cases, the goodness of fit has increased as well, with $\chi^2_\mathrm{min} = 100.0$ from the blinded QSO sample and $\chi^2_\mathrm{min} = 156.5$ from the blinded LRG sample.
In any case, as we can see in figure \ref{fig:pk}, the unblinded power spectra are enhanced around the turnover, making it easier to detect. Our best-fitting values of the turnover scale parameter are $\alpha_\mathrm{TO} = 1.049\pm 0.067$ from the QSO sample and $\alpha_\mathrm{TO} = 0.988\pm 0.047$ from the LRGs, in either case, consistent with $\alpha_\mathrm{TO} = 1$, and,  in turn, with the fiducial cosmological model.

\begin{table}[htbp]
\centering
\begin{tabular}{ccccc}
    \toprule
    \midrule
                   & $m$ & $\mathcal{P}_\mathrm{det}$    & $\alpha_\mathrm{TO}$  & $\frac{\chi^2_\mathrm{min}}{\mathrm{dof}}$ \\
\hline
QSO & $\phantom{+}0.68(0.74)\pm 0.41\phantom{+}$ & 98\% & $\phantom{+}1.049\pm 0.067$ & $\phantom{+}0.92\phantom{+}$\\
LRG & $\phantom{+}0.23(0.28)\pm 0.25\phantom{+}$ & 90\% & $\phantom{+}0.988\pm 0.047$ & $\phantom{+}1.28\phantom{+}$\\
\midrule
    \midrule
    \bottomrule
    \end{tabular}
\caption{Best-fitting parameter values and 1$\sigma$ quantiles for the slope parameter $m$, as well as the turnover dilation parameter $\alpha_\mathrm{TO}$, and the detection probability $\mathcal{P}_\mathrm{det} = \mathcal{P}(m > 0)$. For $m$, we also list the posterior mean in brackets.\label{tab:mcmc_output}}
\end{table}

\subsection{Correlation between Turnover and BAO measurements}

A measurement of $\alpha_\mathrm{TO}$ can be translated into a highly degenerate measurement of $\Omega_\mathrm{m}$ and $H_0$. One way to break this degeneracy while remaining agnostic about the sound-horizon scale $r_\mathrm{d}$ is to incorporate $\Omega_\mathrm{m}$ information from the BAO. In the eBOSS analysis \cite{Bahr-Kalus:eBOSSTO}, we measured the turnover from QSOs only and combined this measurement with BAO measurements from the Lyman-$\alpha$ forest and tracers other than QSOs to avoid correlations between the two datasets. Since we are measuring the turnover from LRGs here, we would also lose a considerable amount of constraining power if we were disregarding the DESI LRG and QSO BAO constraints. 

This is why we use again the EZmock best-fitting values of $\alpha_\mathrm{TO}$ from section \ref{sec:int_crosscorr} and correlate them with the best-fitting BAO parameters $\alpha_\mathrm{iso}$ and $\alpha_\mathrm{AP}$ from the same mocks used in the covariance estimation for the main DESI BAO analysis \cite{DESI2024.III.KP4}. We do not consider the BAO from the DESI Lyman-$\alpha$ analysis \cite{DESI2024.IV.KP6} as it would be more complicated to estimate its correlation with the turnover measurement given that we use QSOs that have been used in the Lyman-$\alpha$ analysis. 

\begin{figure}[htbp]
    \centering
    \includegraphics[width=\columnwidth]{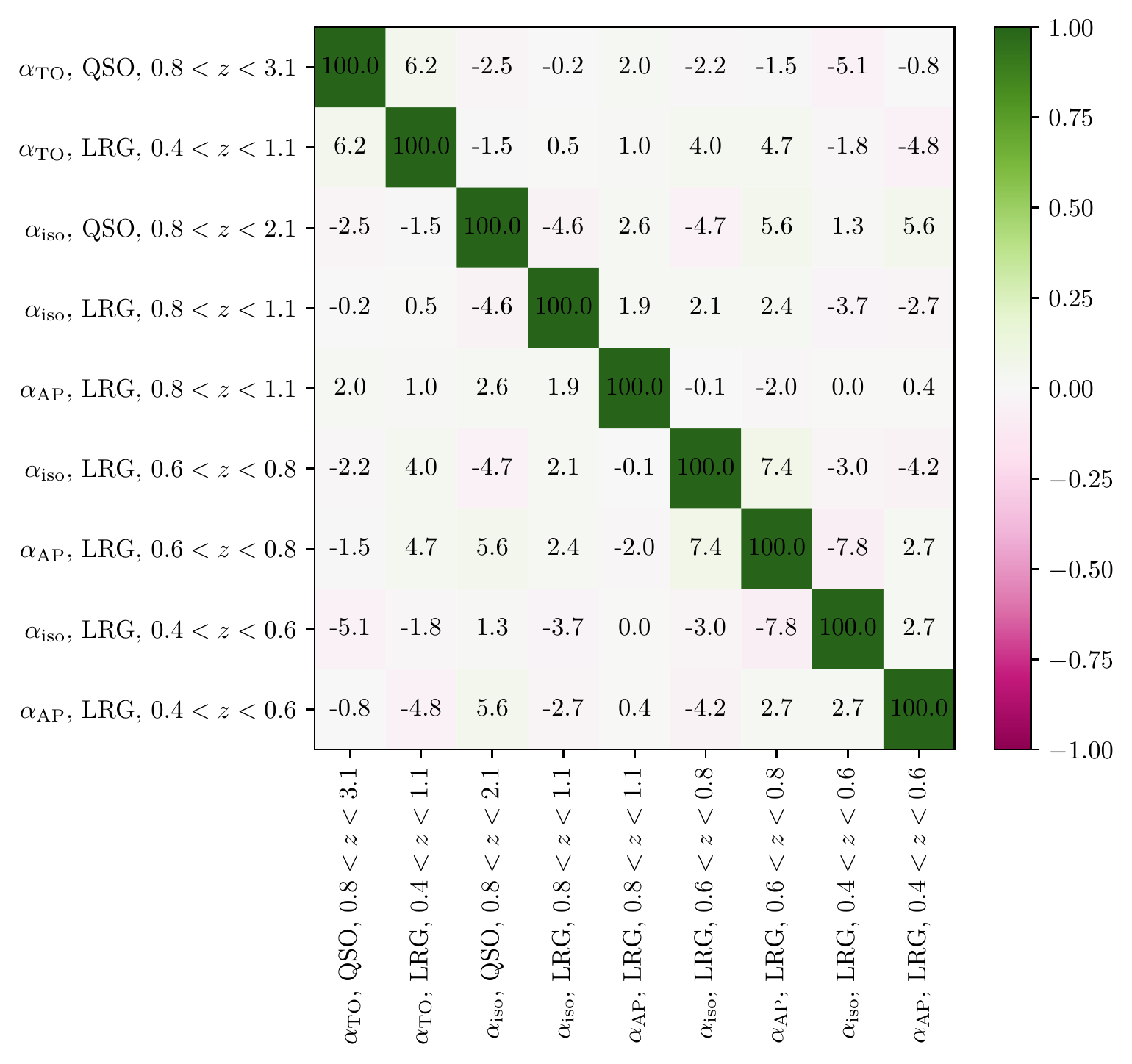}
    \caption{Correlations in percentage between $\alpha_\mathrm{TO}$, as well as post-reconstruction $\alpha_\mathrm{iso}$ and $\alpha_\mathrm{AP}$ for the QSO and LRG samples in per cent. These correlation matrices have been estimated from 1000 EZmocks. We also show the correlations between the QSO and LRG $\alpha_\mathrm{TO}$, as they overlap in redshift.\label{fig:to_bao_corr}}
\end{figure}

We illustrate the turnover-BAO correlations also in figure \ref{fig:to_bao_corr}. All these correlations are weaker than the correlation between the QSO and LRG turnover parameters. The most significant turnover-BAO (anti-)correlation occurs between QSOs and isotropic LRG BAO
in the redshift bin of $0.4<z<0.6$. Since these represent different populations in non-adjacent redshift bins, we consider this and all other less correlated pairs to be coincidental correlations and thus treat them as independent.

Additionally, DESI provides a further model-independent approach to obtain information beyond BAO and RSD parameters, the ShapeFit approach \cite{Brieden:ShapeFit,DESI2024.V.KP5}, which, however, has not been used for cosmological parameter estimation \cite{DESI2024.VII.KP7B}. We investigate the correlation between turnover measurements and ShapeFit parameters in Appendix \ref{app:shapefit_corr}.

\subsection{The Turnover Scale as a Standard Ruler}
\label{sec:TO_as_std_ruler}

\begin{figure}[htbp]
\centering
\includegraphics[width=0.94\columnwidth]{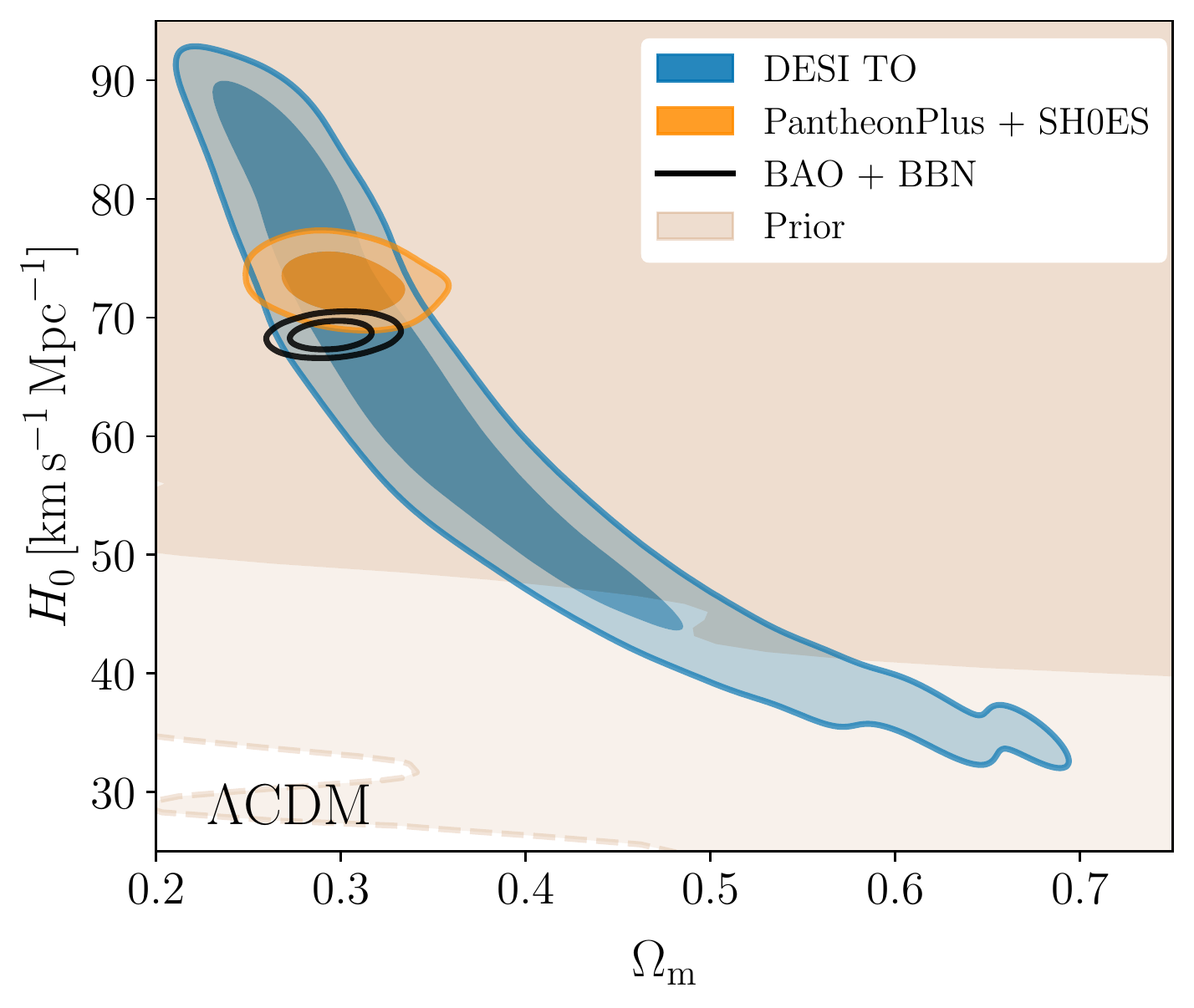}
\includegraphics[width=\columnwidth]{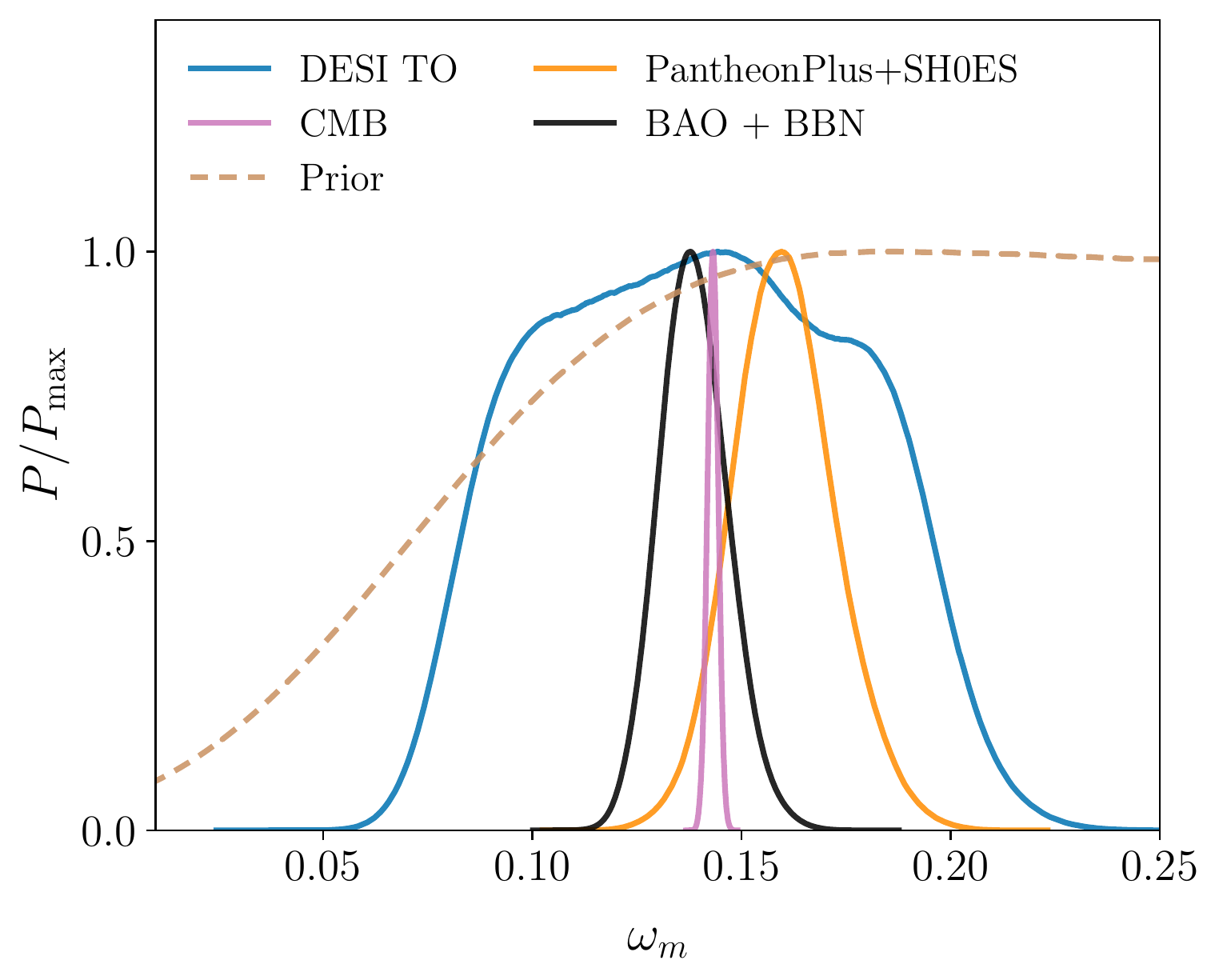}
\caption{Constraints on the $\Lambda$CDM parameters $H_0$ and $\Omega_\mathrm{m}$ (top), as well as $\omega_\mathrm{m}$ (bottom) using DESI turnover measurement (blue). These are compared to constraints from PantheonPlus+SH0ES (yellow), DESI BAO+BBN (black), and Planck (magenta, bottom panel only). Priors are imposed on $\omega_\mathrm{b}$ and $\omega_\mathrm{cdm}$. To illustrate how these priors translate into priors on $H_0$, $\Omega_\mathrm{m}$ and $\omega_\mathrm{m}$, we randomly sample points from the priors and include their contours in gold.}\label{fig:omegamH0_TO_lcdm}
\end{figure}

We have presented a model-independent measurement of the power spectrum turnover scale in table \ref{tab:mcmc_output}. As outlined in section \ref{sec:fidcosmo}, we can use this measurement as a standard ruler to make inferences on a particular cosmological model. We begin with a basic flat $\Lambda$CDM model, wherein \( H(z) \) is defined by \( H_0 \) and \( \Omega_\mathrm{m} \) at lower redshifts, and by \( H_0 \), \( \Omega_\mathrm{r} \), and \( \Omega_\mathrm{m} \) through to the matter-radiation equality epoch. By using equations \eqref{eq:aeq} - \eqref{eq:rH_func_of_omega} and fixing the mean CMB temperature at \( T_\gamma = (2.72548 \pm 0.00057)\; \mathrm{K} \) \citep{Fixsen}, \( \Omega_\mathrm{r} \) effectively becomes dependent solely on \( H_0 \). Consequently, we sample the \( H_0 \)-\( \Omega_\mathrm{m} \) posterior contour represented in {blue} in \autoref{fig:omegamH0_TO_lcdm}. These contours are degenerate, hindering competitive constraints on either \( H_0 \) or \( \Omega_\mathrm{m} \) independently. Nonetheless, as illustrated at the bottom of \autoref{fig:omegamH0_TO_lcdm}, we can sample instead $\omega_\mathrm{m} \equiv \Omega_\mathrm{m}h^2$ to obtain a CMB anisotropy-independent constraint of 
\begin{equation}
\Omega_\mathrm{m}h^2 = {0.139\pm 0.036}.
\end{equation}
This result aligns with \(\Omega_\mathrm{m}h^2 = {0.159^{+0.041}_{-0.037}}\) derived from the eBOSS QSO turnover measurement \cite{Bahr-Kalus:eBOSSTO,Bahr-Kalus:eBOSSTOerr} and \(\Omega_\mathrm{m}h^2 = 0.1430 \pm 0.0011\) obtained from the Planck mission \citep{Planck:params}.

\begin{figure}[htbp]
\centering
\includegraphics[width=\columnwidth]{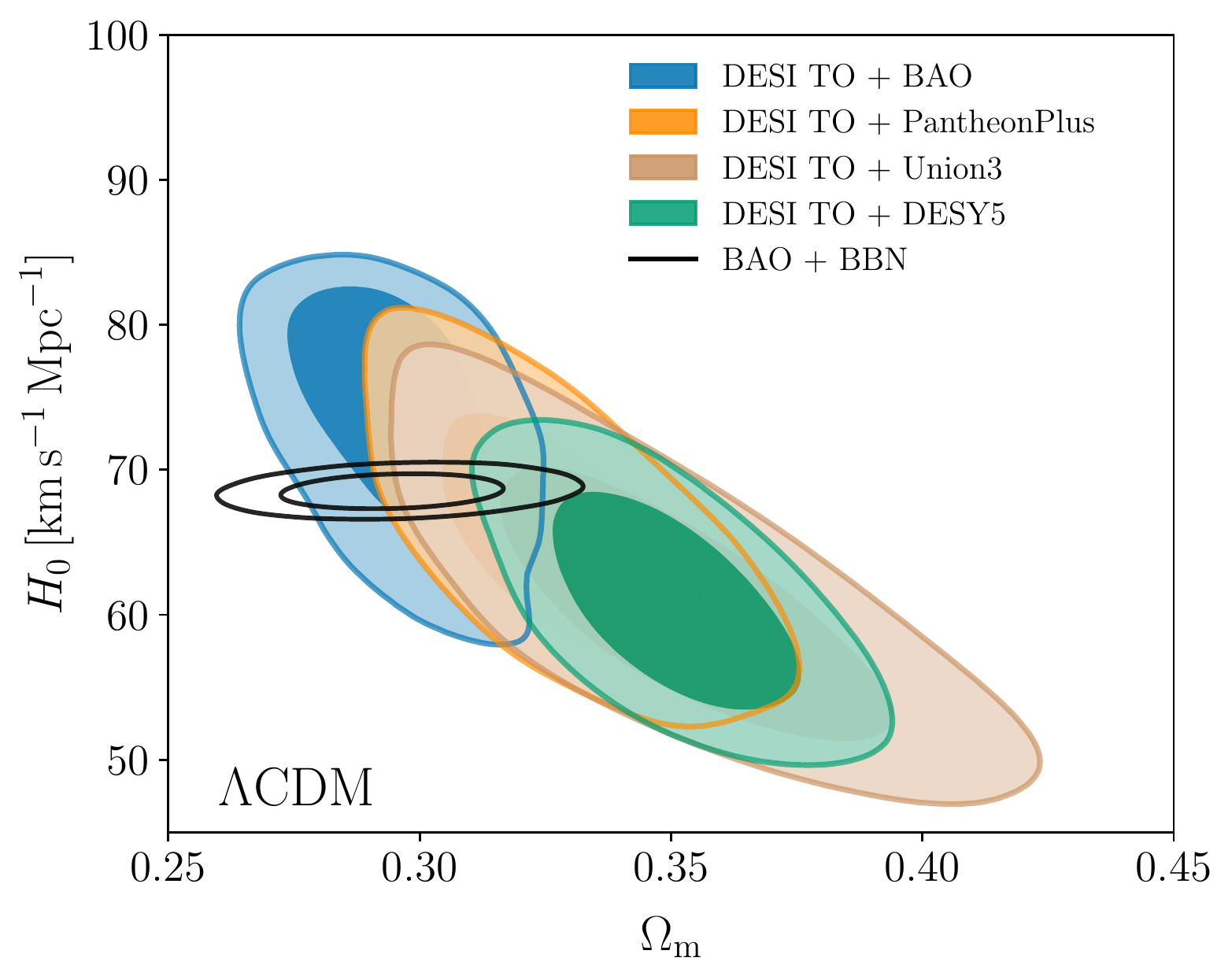}
\includegraphics[width=\columnwidth]{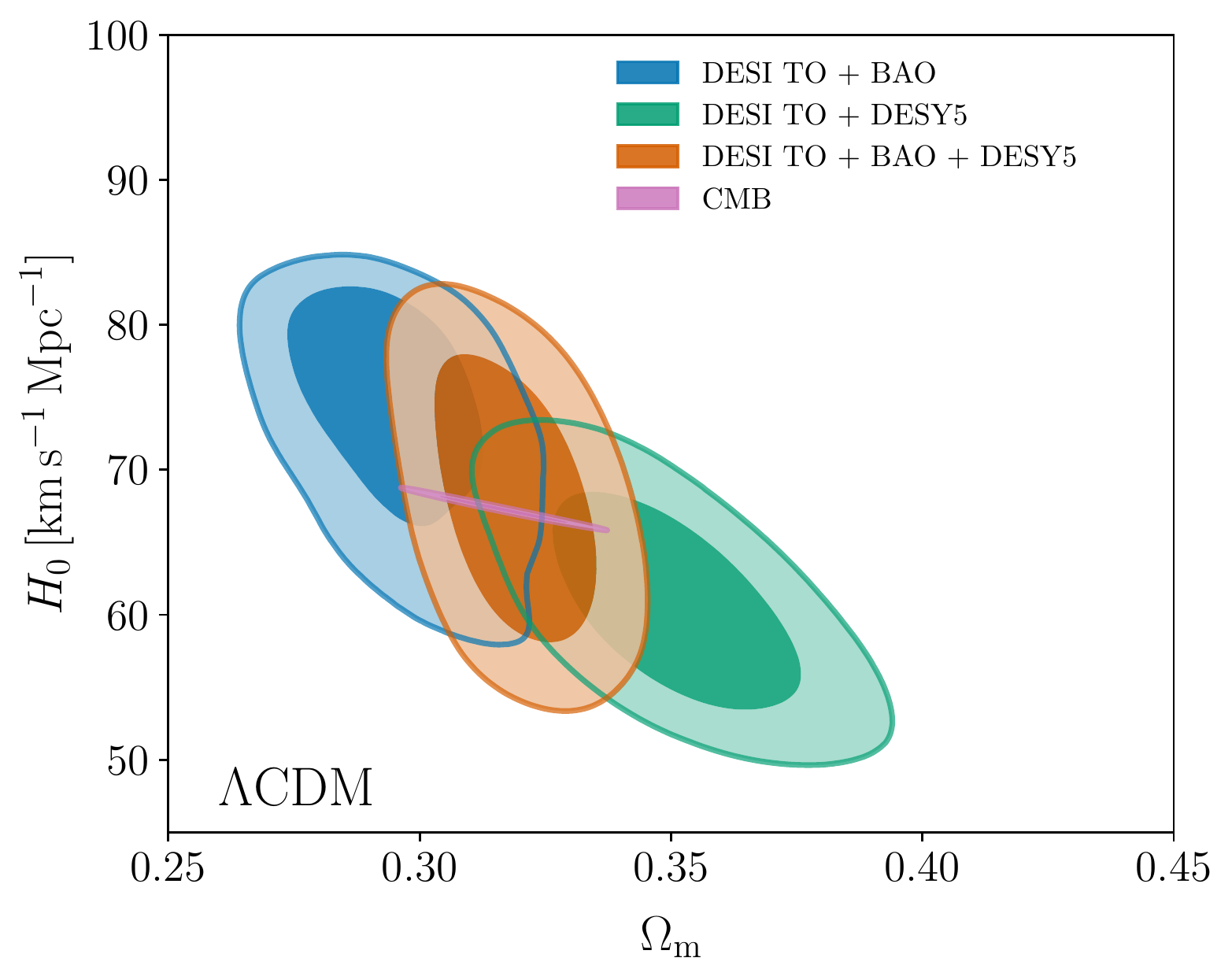}
\caption{\textit{Top:} $\Omega_\mathrm{m}$-$H_0$ posterior contours when combining the DESI turnover measurement with either the DESI BAO, or SN constraints from PantheonPlus \cite{Pantheon:data,Pantheon:params}, Union3 \cite{Union:3}, and DESY5 \cite{DES:SN5YR}. \textit{Bottom:} Similar to top panel exploring combinations of TO, BAO, and SNe and comparing those with CMB constraints.} \label{fig:omegamH0_TO_2data_lcdm}
\end{figure}

\begin{figure*}[htbp]
\centering
\includegraphics[width=0.8\textwidth]{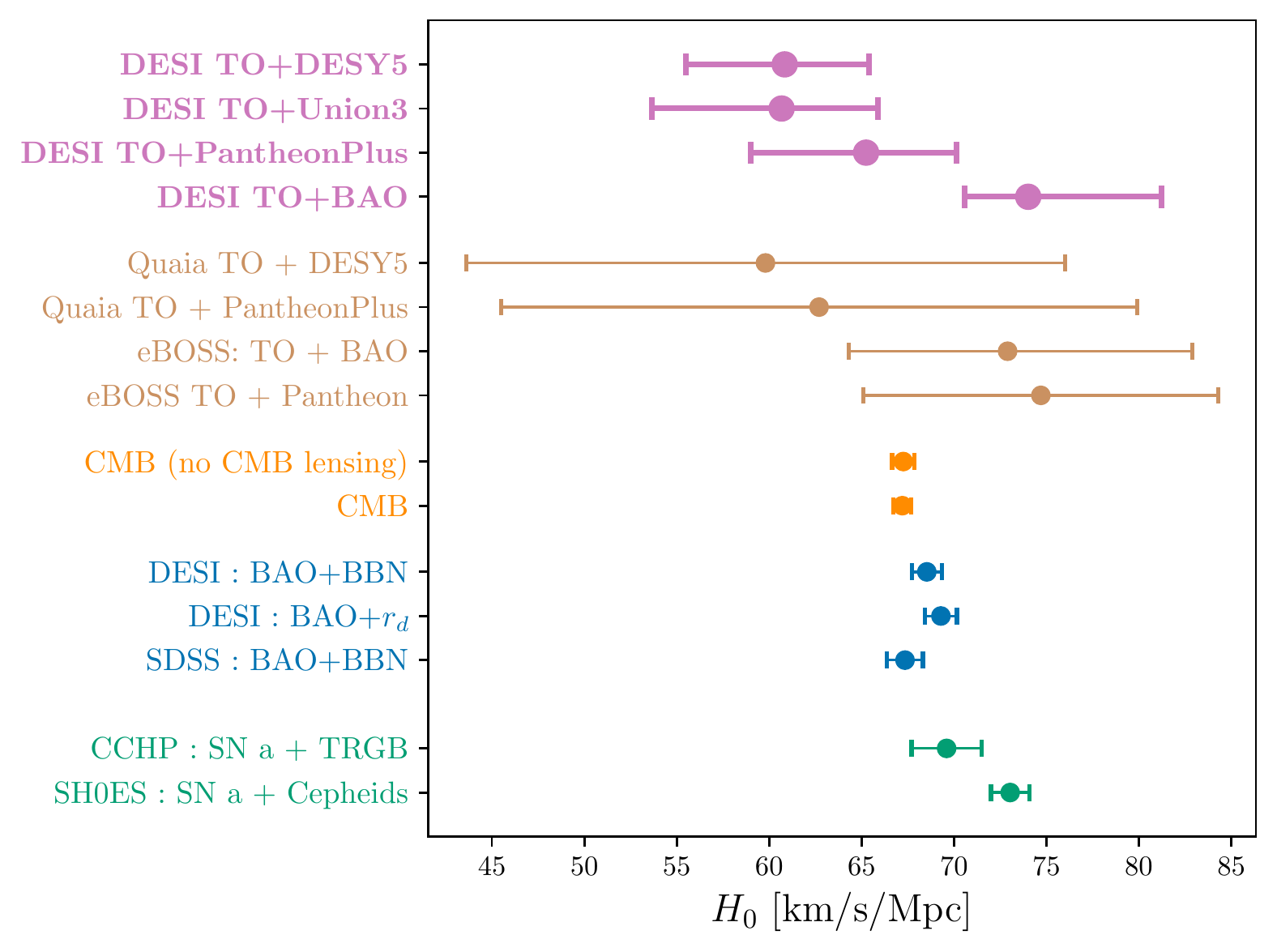}
\caption{Whisker plot comparing the $\Lambda$CDM $H_0$ constraints from DESI TO and the external datasets employed in this work (purple) with other TO based constraints (brown), CMB constraints (orange), BAO constraints obtained with BBN or CMB $r_\mathrm{d}$ priors (blue), and SN distance ladder results (green).}\label{fig:Whisker_H0}
\end{figure*}

\begin{table*}
\centering
\resizebox{\textwidth}{!}{%
    \small
    \begin{tabular}{lcccccc}
    \toprule
    \midrule
    \multirow{2}{*}{model/dataset} & \multirow{2}{*}{$\Omega_{\rm m}$} & $H_0$ & \multirow{2}{*}{$\omega_m$} & \multirow{2}{*}{$w_0$} & \multirow{2}{*}{$w_a$}\\
     & & [$\kmsMpc$] & & & \\
    \midrule
    {\bf Flat} $\boldsymbol{\Lambda}${\bf CDM} &&&&\\
    % DESI & $0.295\pm 0.015$ &---&---&---&---\\
    % DESI+BBN & $0.295\pm 0.015$ & $68.53\pm 0.80$ &---&---&---\\
    % DESI+BBN+$\theta_\ast$ & $0.2948\pm 0.0074$ & $68.52\pm 0.62$ &---&---&---\\
    % DESI+CMB & $0.3069\pm 0.0050$ & $67.97\pm 0.38$ &---&---& --\\
    % \midrule
    % $\boldsymbol{w_0w_a}${\bf CDM} &&&&\\
    % DESI & $0.344^{+0.047}_{-0.026}$ &---&---& $-0.55^{+0.39}_{-0.21}$ & $< -1.32$ \\
    % DESI+BBN+$\theta_\ast$ & $0.338^{+0.039}_{-0.029}$ & $65.0^{+2.3}_{-3.6}$ &---& $-0.53^{+0.42}_{-0.22}$ & $< -1.08$ \\
    % DESI+CMB & $0.344^{+0.032}_{-0.027}$ & $64.7^{+2.2}_{-3.3}$ &---& $-0.45^{+0.34}_{-0.21}$ & $-1.79^{+0.48}_{-1.0}$ \\
    % DESI+CMB+Panth. & $0.3085\pm 0.0068$ & $68.03\pm 0.72$ &---& $-0.827\pm 0.063$ & $-0.75^{+0.29}_{-0.25}$ \\
    % DESI+CMB+Union3 & $0.3230\pm 0.0095$ & $66.53\pm 0.94$ &---& $-0.65\pm 0.10$ & $-1.27^{+0.40}_{-0.34}$ \\
    % DESI+CMB+DESY5 & $0.3160\pm 0.0065$ & $67.24\pm 0.66$ &---& $-0.727\pm 0.067$ & $-1.05^{+0.31}_{-0.27}$ \\
    % \midrule
    DESI TO & $0.354^{+0.030}_{-0.11}$ & $65^{+20}_{-10}$ & $0.139\pm 0.036$ & -- & -- \\
    \midrule
    DESI TO+BAO & $0.295\pm 0.013$ & $74.0^{+7.2}_{-3.5}$ & $0.162^{+0.027}_{-0.017}$ & -- & -- \\
    \midrule
    DESI TO+PantheonPlus & $0.329\pm 0.018$ & $65.2^{+4.9}_{-6.2}$ & $0.140^{+0.017}_{-0.023}$ & -- & -- \\
    \midrule
    DESI TO+Union3 & $0.353\pm 0.026$ & $60.7^{+5.2}_{-7.0}$ & $0.130^{+0.016}_{-0.023}$ & -- & -- \\
    \midrule
    DESI TO+DESY5 & $0.350\pm 0.017$ & $60.8^{+4.6}_{-5.4}$ & $0.130^{+0.016}_{-0.020}$ & -- & -- \\
    \midrule
    DESI TO+BAO+PantheonPlus & $0.308\pm 0.011$ & $70.7^{+7.7}_{-5.9}$ & $0.155\pm 0.026$ & -- & -- \\
    \midrule
    DESI TO+BAO+Union3 & $0.308^{+0.011}_{-0.013}$ & $71.0^{+7.8}_{-5.5}$ & $0.156^{+0.030}_{-0.024}$ & -- & -- \\
    \midrule
    DESI TO+BAO+DESY5 & $0.319\pm 0.011$ & $67.9\pm 6.3$ & $0.148^{+0.024}_{-0.029}$ & -- & -- \\
    \midrule
    CMB & $0.3165\pm 0.0084$ & $67.27\pm 0.60$ & $0.1432\pm 0.0013$ & -- & -- \\
    \midrule
    \midrule
    \textbf{Flat $w_0w_a$CDM} &  &  &  &  &  \\
    \midrule
    DESI TO & $0.374^{+0.060}_{-0.16}$ & $63^{+20}_{-10}$ & $0.140^{+0.021}_{-0.058}$ & $-0.90^{+1.0}_{-0.86}$ & $< -0.560$ \\
    \midrule
    DESI TO+BAO & $0.344^{+0.039}_{-0.029}$ & $66.5\pm 7.2$ & $0.152\pm 0.026$ & $-0.53^{+0.38}_{-0.22}$ & $< -1.32$ \\
    \midrule
    DESI TO+BAO+PantheonPlus & $0.311^{+0.012}_{-0.014}$ & $70^{+8}_{-5}$ & $0.155^{+0.030}_{-0.022}$ & $-0.865\pm 0.073$ & $-0.56^{+0.51}_{-0.39}$ \\
    \midrule
    DESI TO+BAO+Union3 & $0.331\pm 0.015$ & $67^{+8}_{-6}$ & $0.151^{+0.031}_{-0.026}$ & $-0.66\pm 0.12$ & $-1.35^{+0.68}_{-0.59}$ \\
    \midrule
    DESI TO+BAO+DESY5 & $0.324\pm 0.014$ & $68^{+8}_{-6}$ & $0.150^{+0.029}_{-0.027}$ & $-0.738\pm 0.087$ & $-1.11^{+0.57}_{-0.51}$ \\
    \midrule
    BAO + CMB + DESY5 & $0.3163\pm 0.0066$ & $67.23\pm 0.66$ & $0.1429\pm 0.0011$ & $-0.726\pm 0.070$ & $-1.06^{+0.35}_{-0.29}$ \\
    % \\
% \midrule
    \midrule
    \bottomrule
    \end{tabular}
}
\caption{The results for cosmological parameters from turnover measurements, combined with external datasets and priors, are presented within the baseline flat $\Lambda$CDM model and its $w_0 w_a$ extension. We report marginalised means and 68\% credible intervals.
    \vspace{0.1em}
    \label{tab:parameter_table1}
}
\end{table*}

% \begin{table}
% % \midrule
% \begin{tabular}{cccccc}
% \toprule
% \midrule
% model/dataset & $\Omega_\mathrm{m}$ & $H_0$ & $\Omega_m h^2$ & $w_{0,\mathrm{DE}}$ & $w_{a,\mathrm{DE}}$ \\
% \midrule
% \textbf{Flat $\Lambda$CDM} &  &  &  &  &  \\
% \midrule
% TO & $0.354^{+0.030}_{-0.11}$ & $65^{+20}_{-10}$ & $0.139^{+0.036}_{-0.046}$ & -- & -- \\
% \midrule
% TO+BAO & $0.295^{+0.012}_{-0.014}$ & $73.7^{+7.6}_{-3.6}$ & $0.161^{+0.028}_{-0.017}$ & -- & -- \\
% \midrule
% TO+PantheonPlus & $0.330\pm 0.018$ & $64.4^{+5.1}_{-6.2}$ & $0.137^{+0.017}_{-0.022}$ & -- & -- \\
% \midrule
% \textbf{Flat $w_0w_a$CDM} &  &  &  &  &  \\
% \midrule
% TO & $0.367^{+0.055}_{-0.15}$ & $64^{+20}_{-10}$ & $0.142^{+0.023}_{-0.055}$ & $-0.84\pm 0.83$ & $< -0.544$ \\
% \midrule
% TO+BAO+DESY5 & $0.323\pm 0.014$ & $68^{+9}_{-6}$ & $0.148\pm 0.026$ & $-0.743^{+0.081}_{-0.092}$ & $-1.07^{+0.61}_{-0.49}$ \\
% \midrule
% \end{tabular}
% \end{table}

One of the motivations for measuring the turnover was obtaining a sound-horizon-free measurement of the Universe's expansion rate $H_0$. One way to achieve this by relying entirely on DESI data is to break the turnover $H_0$-$\Omega_\mathrm{m}$ degeneracy. This can be done with uncalibrated anisotropic BAO measurements, which provide independent constraints on $\Omega_\mathrm{m}$. We show our joint $\Lambda$CDM $\Omega_\mathrm{m}$-$H_0$ posterior distribution in figure \ref{fig:omegamH0_TO_2data_lcdm} from which we obtain a marginalised mean value of $H_0 = \HOfromTOandBAOBF$. This is almost $2\sigma$ higher than the Planck CMB value of $H_0 = 67.27\pm 0.60$ \cite{Planck:params}. It is also higher than BAO $H_0$ measurements with sound-horizon priors from either Big Bang Nucleosynthesis (BBN) or from the CMB (cf. figure \ref{fig:Whisker_H0}) but it is consistent with the $71.2\pm 4.1\;\mathrm{km/s/Mpc}$-measurement from the DESI galaxy clustering only full-shape analysis where the sound horizon has been marginalised out \cite{DESI:BAO-free}, as well as the eBOSS TO+BAO and eBOSS TO+Pantheon results \cite{Bahr-Kalus:eBOSSTO,Bahr-Kalus:eBOSSTOerr}. 

Our DESI TO+BAO result is also close to the SH0ES measurement of $H_0 = 73.04 \pm 1.04\; \mathrm{km/s/Mpc} $ \cite{Riess:SH0ES} which we also add to the whisker plot in figure \ref{fig:Whisker_H0} for comparison, showing a clear visualisation of how our results compare to previous measurements. As SNe also provide a direct measurement of $\Omega_\mathrm{m}$ but a $H_0$ measurement that is degenerate with intrinsic properties of SNe, we also combine our DESI TO constraints with the SN likelihoods considered in the DESI cosmological parameter inference key articles \cite{DESI2024.VI.KP7A, DESI2024.VII.KP7B}, which are the PantheonPlus\footnote{As in \cite{DESI2024.VII.KP7B}, we denote the originally named Pantheon+ dataset henceforth as PantheonPlus to avoid ambiguity with the '+' symbol used to denote combinations of datasets.} \cite{Pantheon:data,Pantheon:params}, Union3 \cite{Union:3}, and the Year 5 SN analysis from the Dark Energy Survey (DESY5) \cite{DES:SN5YR}. Interestingly, our analysis indicates an inverse Hubble tension. The DESI TO+Union3 and DESI TO+DESY5 combinations yield significantly lower values of \( H_0 \) at \( 60.7^{+5.2}_{-7.0}\; \mathrm{km/s/Mpc}\) and \( 60.8^{+4.6}_{-5.4}\; \mathrm{km/s/Mpc}\), respectively, compared to our DESI TO+BAO constraints (see table \ref{tab:parameter_table1} and figures \ref{fig:omegamH0_TO_2data_lcdm} and \ref{fig:Whisker_H0}). When combining with the PantheonPlus dataset, we observe a milder tension, resulting in a marginalised mean and 68\% credible interval of \( H_0 = 65.2^{+4.9}_{-6.2} \; \mathrm{km/s/Mpc} \). It is important to note that this 'inverse Hubble tension' is not indicative of an inconsistency within the DESI TO constraints. Instead, it reflects the differences in the values of \( \Omega_\mathrm{m} \): \( 0.295 \pm 0.015 \) from DESI BAO \cite{DESI2024.VI.KP7A} versus \( 0.356^{+0.028}_{-0.026} \) from Union3 \cite{Union:3}, to give as an example the datasets that result in the highest tension when combining with the DESI TO.

\begin{figure*}[htbp]
\centering
\includegraphics[width=\textwidth]{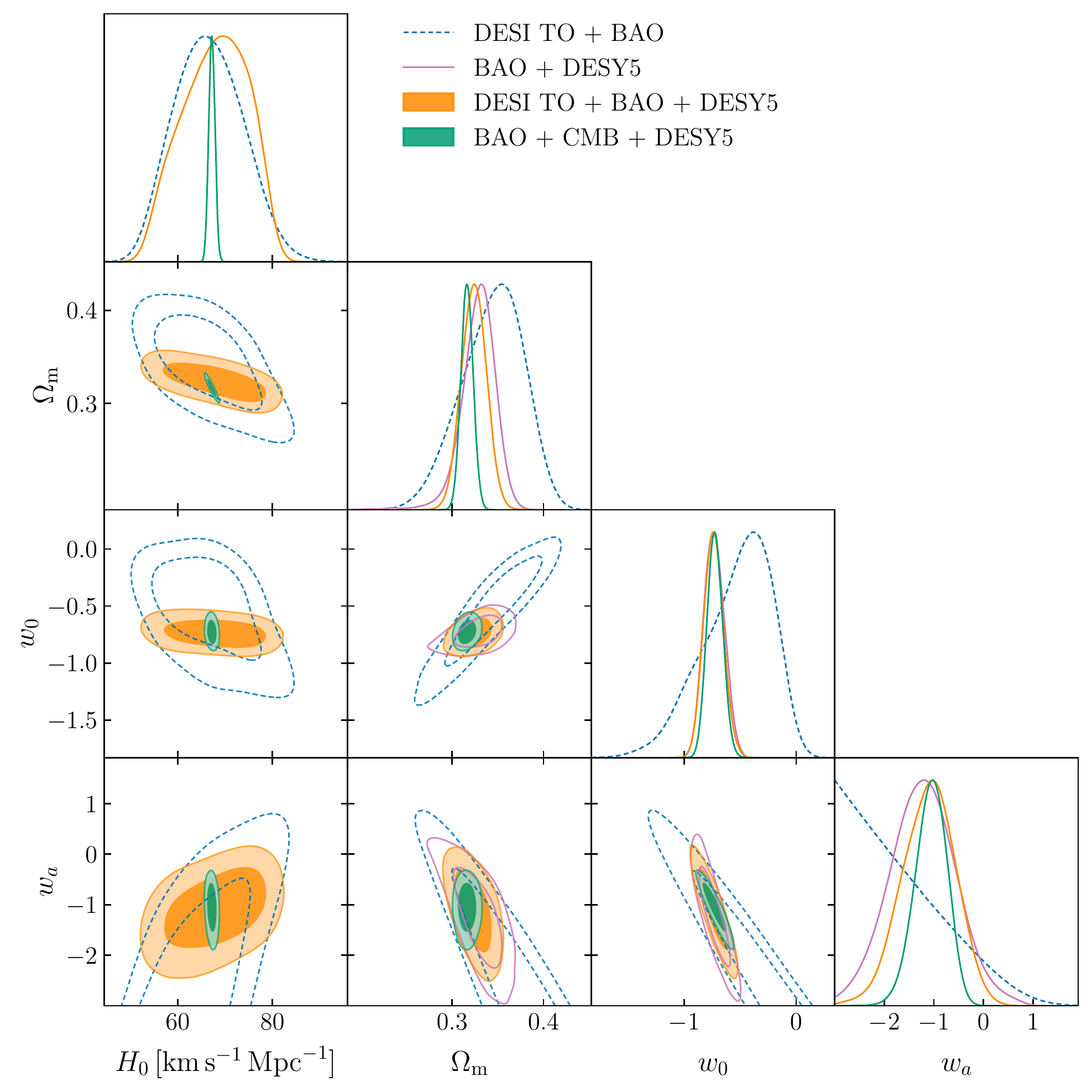}
\caption{Posterior contours for $H_0$, $\Omega_\mathrm{m}$, and the dark energy parameters $w_0$ and $w_a$. Note, that we do not show any $H_0$ contours for BAO + DESY5 as these are $r_\mathrm{d}$ dependent \cite{DESI2024.VI.KP7A}. 
}\label{fig:omegamH0_TO_2data_wcdm}
\end{figure*}

This `$\Omega_\mathrm{m}$ tension' can be resolved by allowing for evolving dark energy. In this model, DESI BAO constrains the matter density parameter at $\Omega_\mathrm{m} = 0.344^{+0.047}_{-0.026}$ \cite{DESI2024.VI.KP7A}. In fact, DESI BAO+CMB+SNe favour evolving dark energy over $\Lambda$CDM at the $2.5\sigma$, $3.5\sigma$ or $3.9\sigma$ level depending on whether the SN sample is PantheonPlus, Union3, or DESY5 \cite{DESI2024.VI.KP7A}. Adopting the Chevallier-Polarski-Linder (CPL, \cite{Chevallier:CPL,Linder:CPL}) parametrisation where the dark energy equation of state $w(a) = w_0 + w_a(1 - a)$ at scale factor $a$ is defined by the present-day value $w_0$ and the linear evolution parameter $w_a$, we can see that in this $w_0w_a$CDM model, DESI TO+BAO provides an $H_0$ constraint of $66.5 \pm 7.2\;\mathrm{km/s/Mpc}$ that agrees well with DESI TO+SNe. 

Furthermore, we compare in figure \ref{fig:omegamH0_TO_2data_wcdm} the posterior contours of the data combination that provided the strongest evidence for evolving dark energy, i.e. DESI BAO+DESY5+CMB, to that from replacing the CMB with the DESI TO. While the TO yields less information than the CMB, it still provides a valuable consistency check when combined with BAO and supernovae but without directly relying on the CMB. The TO+BAO+DESY5 combination achieves competitive constraints with 
$w_0 = -0.738\pm 0.087$ and $w_a = -1.11_{-0.51}^{+0.57}$, which are not far from the full DESI+CMB+DESY5 combination ($w_0 = -0.726 \pm 0.070$, $w_a = -1.06_{-0.29}^{+0.35}$, cf. figure \ref{fig:omegamH0_TO_2data_wcdm}). This suggests that the turnover can serve as an independent low-redshift probe, offering cross-checks and robustness tests against CMB-driven cosmological models.

\section{Summary}
\label{sec:summary}

In this study, we present the first model-independent detection of the power spectrum turnover in an auto-power spectrum using data obtained from DESI during its inaugural survey year, with a detection probability of 98 per cent in the QSO sample and 90 per cent in the LRG sample. The power spectrum turnover serves as a model-independent measurement of the matter-radiation equality scale. Our analysis, conducted with strict data blinding techniques to mitigate confirmation bias, yields $D_\mathrm{V}(z = 1.651) = \left(38.1\pm 2.5\right)r_\mathrm{H}$ from the quasars and $D_\mathrm{V}(z = 0.733) = \left(21.8\pm 1.0\right)r_\mathrm{H}$ from the LRG sample. These turnover scales are a crucial feature in the matter power spectrum and serve as a standard ruler in cosmology, complementary to the BAO. 

Assuming three standard neutrino species and a vanilla $\Lambda$CDM cosmology, these measurements allow us to derive a constraint on the matter density parameter \( \Omega_\mathrm{m}h^2 = 0.139^{+0.036}_{-0.046} \). By combining this with low-redshift measurements from type-Ia supernovae data (PantheonPlus, Union3, and DESY5), we obtain an estimate of the Hubble-Lema\^itre parameter independent from BBN or CMB priors: \( H_0 = 60.7^{+5.2}_{-7.0}\; \mathrm{km/s/Mpc}\), \( 60.8^{+4.6}_{-5.4}\; \mathrm{km/s/Mpc}\), and \(\HOfromTOandSNeBF \), respectively, which aligns with previous Quaia+Planck turnover analyses.

Furthermore, incorporating DESI BAO results, we estimate \( H_0 = \HOfromTOandBAOBF \), consistent with findings from DESI full-shape analyses that marginalise the sound-horizon scale \cite{DESI:BAO-free}. Note that this value presents true constraints from DESI alone without any external data, whereas $H_0$ constraints from the BAO alone are either presented as $H_0r_\mathrm{d}$ or are broken with external data, such as in figure \ref{fig:omegamH0_TO_lcdm} where the $H_0$-$r_\mathrm{d}$ degeneracy has been broken with external BBN priors. The apparent inversion of $H_0$ values between figure \ref{fig:omegamH0_TO_lcdm} on one hand and figures \ref{fig:omegamH0_TO_2data_lcdm} and \ref{fig:Whisker_H0} on the other, is due to the exclusion of priors from BBN and SH0ES Cepheid host distances.

Notably, the (now inverted) discrepancy in \( H_0 \) arises due to differences in the matter density measured by supernovae compared to those measured by the DESI BAO. When allowing for evolving dark energy, these differences are reconciled with DESI TO+BAO providing $H_0 = 66.5\pm 7.2\;\mathrm{km/s/Mpc}$. However, the constraints on \( H_0 \) become less stringent. This work underscores the capacity of the turnover scale as a robust cosmological probe, providing valuable insights into the dynamics of the Universe. 

Importantly, this analysis only uses data from the first year of DESI, representing just half of the area of the expected final sample. With four more years of observations planned, the final DESI dataset will contain approximately three times the effective volume of the first year data \cite{DESI2024.III.KP4}, significantly enhancing statistical precision and constraining power. Additionally, there are substantial efforts in addressing emission line galaxy (ELG) systematics, which have prohibited including them in this analysis. The recent implementation of improved imaging calibrations will reduce systematic biases. These improvements pave the way for incorporating ELG samples into future turnover analyses, increasing the number of available tracers by 50 per cent, and, in turn, establishing the power spectrum turnover as a competitive additional probe. 

\section{Data Availability}

Data from the plots in this paper are available on Zenodo at \cite{bahr_kalus_2025_17018002} as part of DESI's Data Management Plan.  The data used in this analysis has been made public along with Data Release 1 of DESI \cite{2025arXiv250314745D}.\footnote{Details can be found here: \url{https://data.desi.lbl.gov/doc/releases/}}

%%%%%%%%%%%%%%%%% ACKNOWLEDGEMENTS %%%%%%%%%%%%%%%%%%%%%
\acknowledgments

BB-K thanks Matilde Barberi-Squarotti for her help with \texttt{getdist}.

BB-K acknowledges support from INAF for the project
``Paving the way to radio cosmology in the SKA Observatory era: synergies between SKA pathfinders/precursors
and the new generation of optical/near-infrared cosmological surveys'' (CUP C54I19001050001).

This research was conducted using multiple high-performance computing resources. In addition to the facilities acknowledged by the Dark Energy Spectroscopic Instrument (DESI) and Legacy Surveys, including the National Energy Research Scientific Computing Center (NERSC), a Department of Energy User Facility, some of the cosmological analysis for this specific work was supported by the Seondeok high-performance computing clusters at the Korea Astronomy and Space Science Institute. Finally, to address specific comments from the referee, this work also made use of Pleiadi \citep{2020ASPC..527..307T,2020ASPC..527..303B}, a computing infrastructure installed and managed by INAF-USCVIII, under the proposal `Simulations and Analyses Across the Spectrum - Setting the Stage for Euclid-SKAO Synergies'.

This material is based upon work supported by the U.S.\ Department of Energy (DOE), Office of Science, Office of High-Energy Physics, under Contract No.\ DE-AC02-05CH11231, and by the National Energy Research Scientific Computing Center, a DOE Office of Science User Facility under the same contract. Additional support for DESI was provided by the U.S. National Science Foundation (NSF), Division of Astronomical Sciences under Contract No.\ AST-0950945 to the NSF National Optical-Infrared Astronomy Research Laboratory; the Science and Technology Facilities Council of the United Kingdom; the Gordon and Betty Moore Foundation; the Heising-Simons Foundation; the French Alternative Energies and Atomic Energy Commission (CEA); the National Council of Humanities, Science and Technology of Mexico (CONAHCYT); the Ministry of Science and Innovation of Spain (MICINN), and by the DESI Member Institutions: \url{https://www.desi. lbl.gov/collaborating-institutions}. 

The DESI Legacy Imaging Surveys consist of three individual and complementary projects: the Dark Energy Camera Legacy Survey (DECaLS), the Beijing-Arizona Sky Survey (BASS), and the Mayall z-band Legacy Survey (MzLS). DECaLS, BASS and MzLS together include data obtained, respectively, at the Blanco telescope, Cerro Tololo Inter-American Observatory, NSF NOIRLab; the Bok telescope, Steward Observatory, University of Arizona; and the Mayall telescope, Kitt Peak National Observatory, NOIRLab. NOIRLab is operated by the Association of Universities for Research in Astronomy (AURA) under a cooperative agreement with the National Science Foundation. Pipeline processing and analyses of the data were supported by NOIRLab and the Lawrence Berkeley National Laboratory. Legacy Surveys also uses data products from the Near-Earth Object Wide-field Infrared Survey Explorer (NEOWISE), a project of the Jet Propulsion Laboratory/California Institute of Technology, funded by the National Aeronautics and Space Administration. Legacy Surveys was supported by: the Director, Office of Science, Office of High Energy Physics of the U.S. Department of Energy; the National Energy Research Scientific Computing Center, a DOE Office of Science User Facility; the U.S. National Science Foundation, Division of Astronomical Sciences; the National Astronomical Observatories of China, the Chinese Academy of Sciences and the Chinese National Natural Science Foundation. LBNL is managed by the Regents of the University of California under contract to the U.S. Department of Energy. The complete acknowledgments can be found at \url{https://www.legacysurvey.org/}.

Any opinions, findings, and conclusions or recommendations expressed in this material are those of the author(s) and do not necessarily reflect the views of the U.S.\ National Science Foundation, the U.S.\ Department of Energy, or any of the listed funding agencies.

The authors are honored to be permitted to conduct scientific research on Iolkam Du'ag (Kitt Peak), a mountain with particular significance to the Tohono O'odham Nation.

\appendix
\section{Test of $\alpha_\mathrm{TO}$ scaling relation on mocks}
\label{app:alpha_scaling}
\begin{figure}[htbp]
\centering
    \includegraphics[width=\columnwidth]{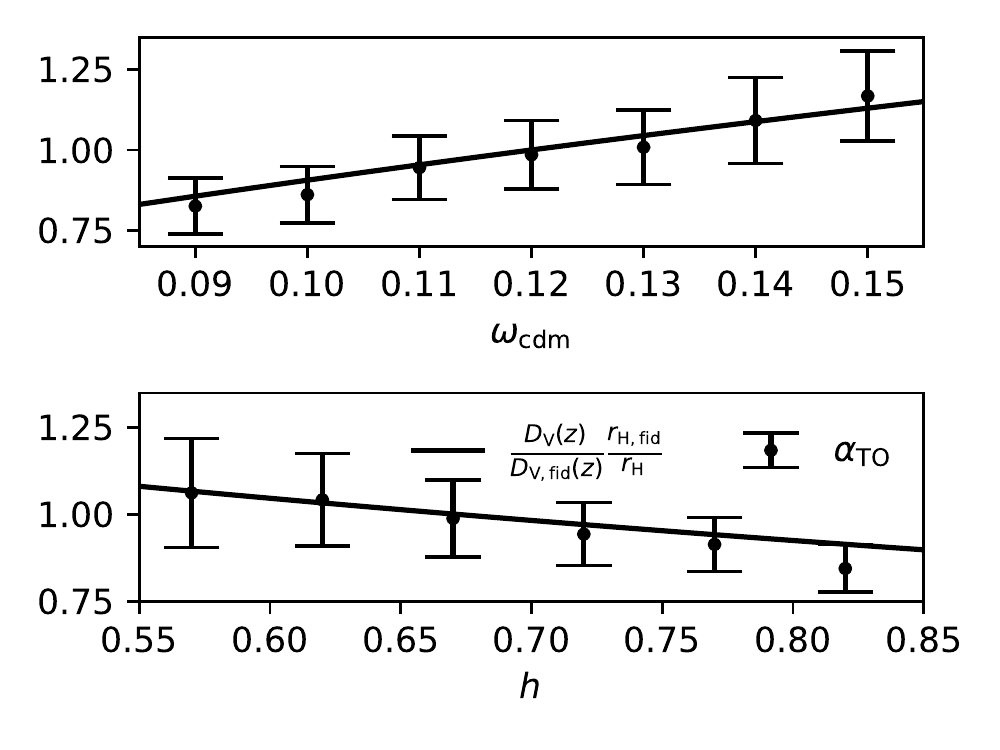}
\caption{Test of the $\alpha_\mathrm{TO}$ scaling relation.  Shown is the relative difference of the left- and right-hand sides of Eq.~\eqref{eq:alphaeq} as a function of $\omega_\mathrm{cdm}$ and $h$. The data points correspond to $\alpha_\mathrm{TO}$ from simulations, whereas the solid lines show the expectation from the right-hand side of equation \eqref{eq:alphaeq}. This demonstrates the validity of the scaling relation for a range of cosmological parameters. \label{fig:alphaTO_scaling}}
\end{figure}

This appendix details the test of the scaling relation for $\alpha_\mathrm{TO}$ given in Eq.~\eqref{eq:alphaeq} using mock galaxy catalogues.  Figure~\ref{fig:alphaTO_scaling} shows the residuals of this scaling relation as a function of cosmological parameters.  The test involves generating mock data, computing the power spectrum, and fitting for the turnover scale. The use of mocks is necessary to test the impact of the fiducial cosmology.

\begin{enumerate}
    \item \textbf{Mock Data Generation:}
    Mock galaxy catalogues were generated using \texttt{mockfactory}'s\footnote{\url{https://github.com/cosmodesi/mockfactory}} \texttt{LagrangianLinearMock} routine, which first creates a linear density field with a DESI-like fiducial cosmology implemented in \texttt{cosmoprimo}. Then, it displaces particles from their Lagrangian grid positions using the first-order Lagrangian perturbation theory (LPT) -- i.e., the Zel'dovich approximation.  The fiducial parameters are:
    \begin{itemize}
        \item Redshift: $z = 1.5$
        \item Linear power spectrum from the DESI fiducial model
        \item Number density: $\bar{n} = 10^{-3}h^3/\mathrm{Mpc}^3$ % Corrected units
        \item Simulation box parameters: \texttt{boxsize = 3500 Mpc/h, nmesh = 256}
    \end{itemize}
    Poisson sampling was used to populate the density field with galaxies.

    \item \textbf{Power Spectrum Computation:}
    The power spectrum monopole was computed using \texttt{pypower.CatalogFFTPower}.  Shot noise and normalization were accounted for. The covariance matrix was estimated assuming a Gaussian approximation:
    \[ C(k_i, k_j) = \left[\frac{2}{V_n V_{\text{eff}}} P^2(k_i) + s_{n,0}^2\right]\delta^\mathrm{K}_{ij} \]
    where \(V_n\) are the Fourier-space volume elements.  The effective volume, $V_\mathrm{eff}$, was calculated consistently with the main analysis.

    \item \textbf{Turnover Scale Fitting:}
    The turnover scale, $\alpha_\mathrm{TO}$, was extracted from the mock power spectra using the same fitting pipeline described in the main text.  Briefly, this involves fitting a model to the power spectrum around the turnover feature.

    \item \textbf{Testing the Scaling Relation:}
    To assess the validity of the scaling relation in Eq.~\eqref{eq:alphaeq}, we varied the values of $\omega_\mathrm{cdm}$ and $h$ around the fiducial cosmology. For each variation:
    \begin{itemize}
        \item The fiducial cosmology used in Eq.~\eqref{eq:alphaeq} and when computing $r_\mathrm{H,fid}$ remained fixed. 
        \item The mock power spectrum was recomputed using the updated cosmology.
        \item The turnover scale, $\alpha_\mathrm{TO}$, was re-fitted.
        \item $\alpha_\mathrm{TO}$ was then compared to its prediction from Eq.~\eqref{eq:alphaeq}, using $r_\mathrm{H}$ calculated for the varied cosmology.
    \end{itemize}
    The datapoints shown in Figure~\ref{fig:alphaTO_scaling} represent the left-hand side of equation \eqref{eq:alphaeq}, whereas the solid lines show the expectation from the right-hand side for these varied cosmologies.
\end{enumerate}

\section{Correlation with ShapeFit}
\label{app:shapefit_corr}
Here, we analyse the correlation of the turnover parameters with ShapeFit parameters. ShapeFit extends the standard BAO and RSD measurements by including a parameter $m_\mathrm{SF}$, which characterises the shape of the power spectrum, and $df$, which accounts for deviations from standard growth predictions \cite{Brieden:ShapeFit}.

\begin{figure}[htbp]
    \centering
    \includegraphics[width=0.47\textwidth]{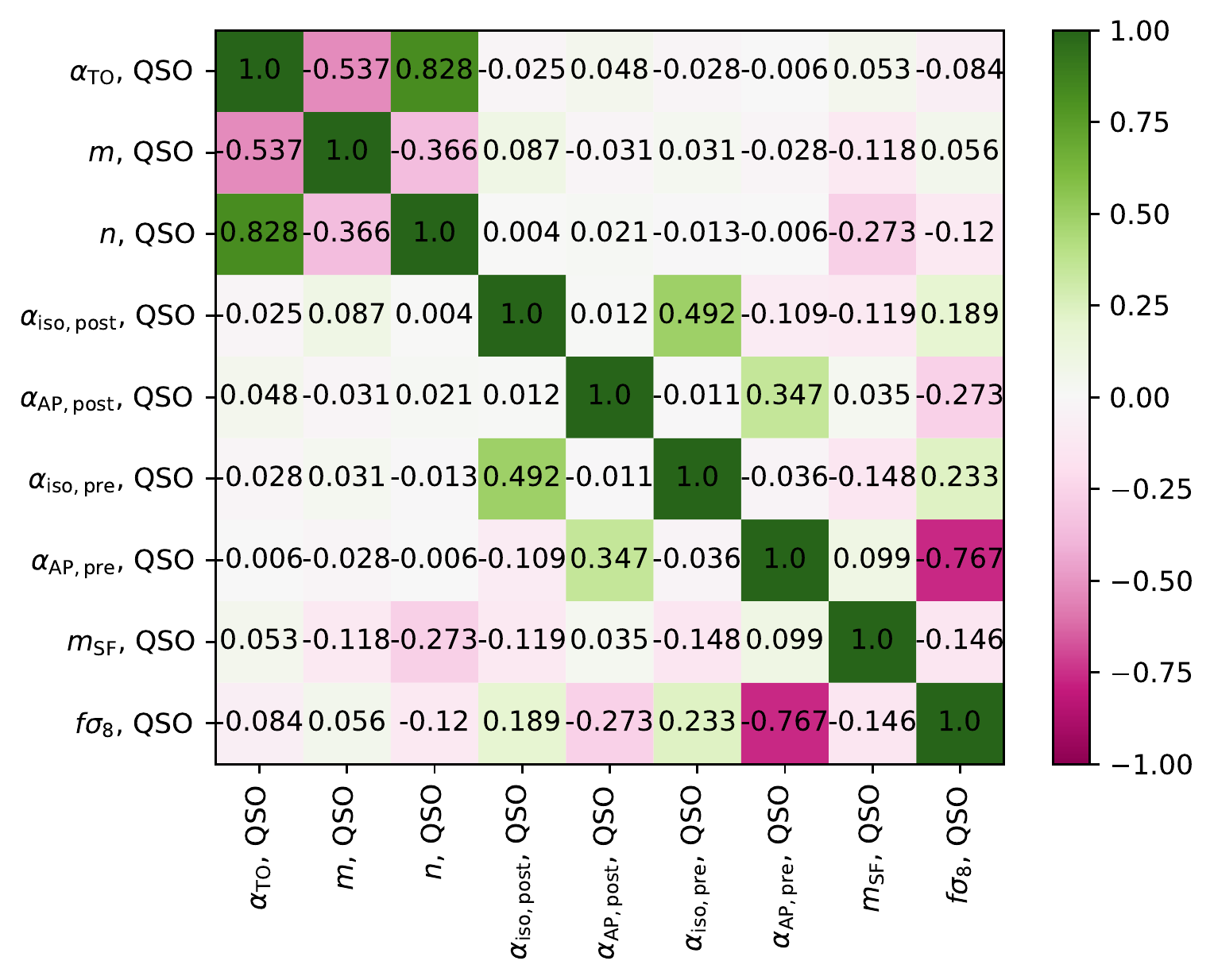}
    \qquad
    \includegraphics[width=\columnwidth]{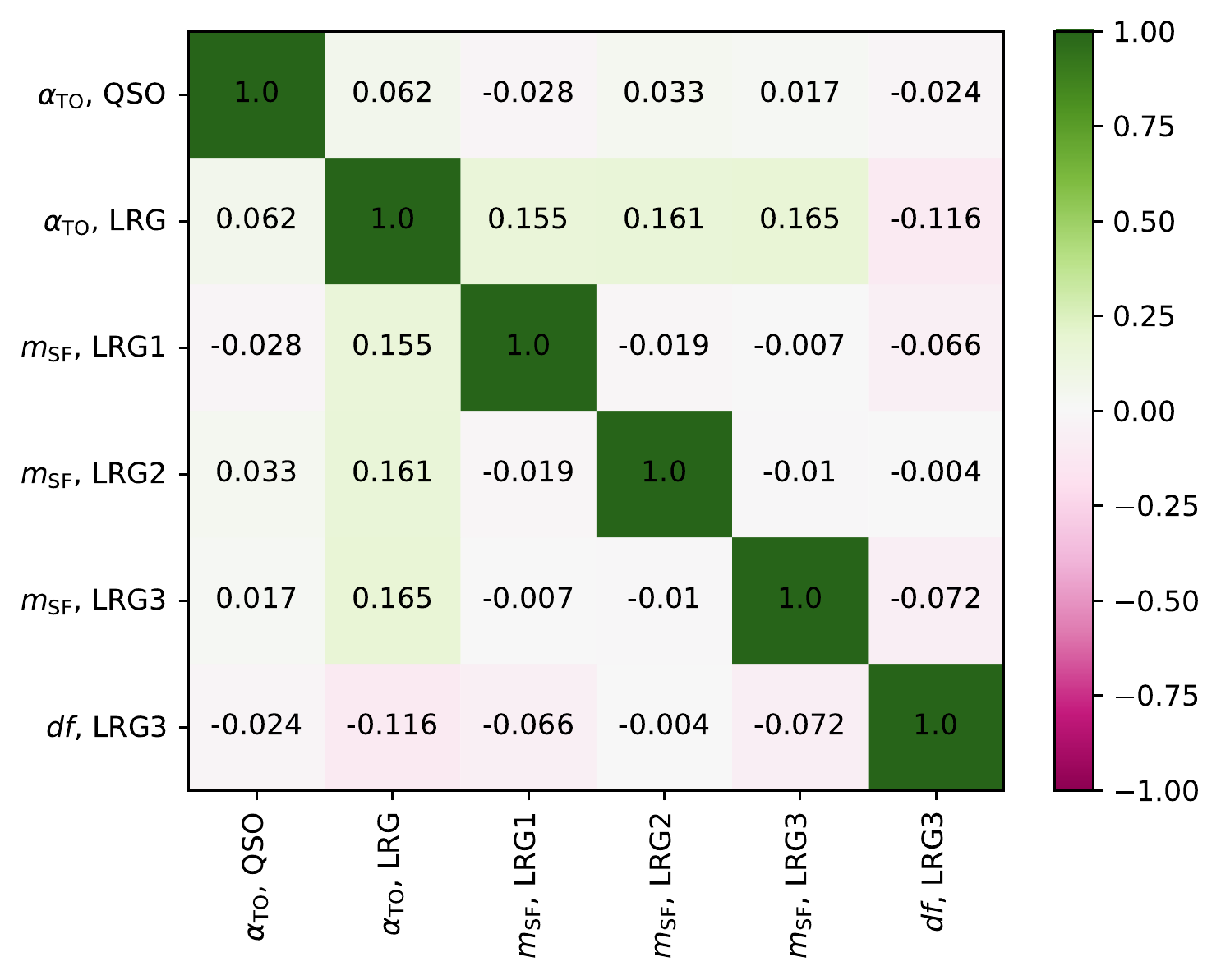}
    \caption{Correlation matrices between $\alpha_\mathrm{TO}$, $\alpha_\mathrm{iso}$ and $\alpha_\mathrm{AP}$ pre- and post-reconstruction, as well as the ShapeFit parameter $m_\mathrm{SF}$ and $df$ values for the QSO (left) and LRG (right) samples. These correlation matrices have been estimated from 1000 EZmocks. As the LRG sample is split into three redshifts bins for the BAO and ShapeFit analyses, we refer with LRG1, LRG2, and LRG3 to the redshift bins with $0.4<z<0.6$, $0.6<z<0.8$, and $0.8<z<1.1$, respectively. We also show the correlations between the QSO and LRG $\alpha_\mathrm{TO}$, as they overlap in redshift. For clarity, in the right-hand panel, we only show parameters whose correlation with the LRG $\alpha_\mathrm{TO}$ is larger than 0.1.\label{fig:to_bao_sf_corr}}
\end{figure}

Using 1000 EZmock realizations, we estimate the correlation matrices between $\alpha_\mathrm{TO}$, $\alpha_\mathrm{iso}$, and $\alpha_\mathrm{AP}$ pre- and post-reconstruction, as well as $m_\mathrm{SF}$ and $df$ for the QSO and LRG samples. These correlations are displayed in figure \ref{fig:to_bao_sf_corr}.

Since the LRG sample is split into three redshift bins for the BAO and ShapeFit analyses, we refer to these bins as LRG1 ($0.4<z<0.6$), LRG2 ($0.6<z<0.8$), and LRG3 ($0.8<z<1.1$). We also show the correlations between the QSO and LRG $\alpha_\mathrm{TO}$ values due to their overlapping redshift range.

Overall, we find that the correlation between turnover and ShapeFit parameters is moderate, with the most significant correlations occurring between $n$ and $m_\mathrm{SF}$ which is expected due to both parameters describing the power spectrum shape at scales smaller than the turnover. In terms of $\alpha_\mathrm{TO}$, we do not see any strong correlations with ShapeFit parameters. This analysis suggests that incorporating ShapeFit parameters into joint fits may provide additional cosmological insights beyond standard BAO and turnover measurements since the shape parameter $n$ is treated as a nuisance parameter in this analysis, whereas the related ShapeFit parameter $m_\mathrm{SF}$ is what provides additional information in the ShapeFit approach. It, therefore, seems worthwhile future work to combine the two parameterisations into a joint one, taking the information of both the turnover position and the slope of the power spectrum close to it into account.

% Bibliography

%% [A] Recommended: using JHEP.bst file
\bibliographystyle{JHEP}
\bibliography{biblio.bib}

\providecommand{\href}[2]{#2}\begingroup\raggedright\begin{thebibliography}{10}

\bibitem{Poole:WiggleZTO}
G.B.~{Poole}, C.~{Blake}, D.~{Parkinson}, S.~{Brough}, M.~{Colless},
  C.~{Contreras} et~al., \emph{{The WiggleZ Dark Energy Survey: probing the
  epoch of radiation domination using large-scale structure}},
  \href{https://doi.org/10.1093/mnras/sts431}{\emph{\mnras} {\bfseries 429}
  (2013) 1902} [\href{https://arxiv.org/abs/1211.5605}{{\ttfamily 1211.5605}}].

\bibitem{Bahr-Kalus:eBOSSTO}
B.~{Bahr-Kalus}, D.~{Parkinson} and E.-M.~{Mueller}, \emph{{Measurement of the
  matter-radiation equality scale using the extended baryon oscillation
  spectroscopic survey quasar sample}},
  \href{https://doi.org/10.1093/mnras/stad1867}{\emph{\mnras} {\bfseries 524}
  (2023) 2463} [\href{https://arxiv.org/abs/2302.07484}{{\ttfamily
  2302.07484}}].

\bibitem{Alonso:TO}
D.~{Alonso}, O.~{Hetmantsev}, G.~{Fabbian}, A.~{Slosar} and K.~{Storey-Fisher},
  \emph{{Measurement of the power spectrum turnover scale from the
  cross-correlation between CMB lensing and Quaia}},
  \href{https://doi.org/10.48550/arXiv.2410.24134}{\emph{arXiv e-prints} (2024)
  arXiv:2410.24134} [\href{https://arxiv.org/abs/2410.24134}{{\ttfamily
  2410.24134}}].

\bibitem{DESI2016b.Instr}
{DESI Collaboration}, A.~{Aghamousa}, J.~{Aguilar}, S.~{Ahlen}, S.~{Alam},
  L.E.~{Allen} et~al., \emph{{The DESI Experiment Part II: Instrument Design}},
  {\emph{arXiv e-prints} (2016) arXiv:1611.00037}
  [\href{https://arxiv.org/abs/1611.00037}{{\ttfamily 1611.00037}}].

\bibitem{FocalPlane.Silber.2023}
J.H.~{Silber}, P.~{Fagrelius}, K.~{Fanning}, M.~{Schubnell}, J.N.~{Aguilar},
  S.~{Ahlen} et~al., \emph{{The Robotic Multiobject Focal Plane System of the
  Dark Energy Spectroscopic Instrument (DESI)}},
  \href{https://doi.org/10.3847/1538-3881/ac9ab1}{\emph{\aj} {\bfseries 165}
  (2023) 9} [\href{https://arxiv.org/abs/2205.09014}{{\ttfamily 2205.09014}}].

\bibitem{Corrector.Miller.2023}
T.N.~{Miller}, P.~{Doel}, G.~{Gutierrez}, R.~{Besuner}, D.~{Brooks}, G.~{Gallo}
  et~al., \emph{{The Optical Corrector for the Dark Energy Spectroscopic
  Instrument}}, \href{https://doi.org/10.3847/1538-3881/ad45fe}{\emph{\aj}
  {\bfseries 168} (2024) 95}
  [\href{https://arxiv.org/abs/2306.06310}{{\ttfamily 2306.06310}}].

\bibitem{FiberSystem.Poppett.2024}
C.~{Poppett}, L.~{Tyas}, J.~{Aguilar}, C.~{Bebek}, D.~{Bramall}, T.~{Claybaugh}
  et~al., \emph{{Overview of the Fiber System for the Dark Energy Spectroscopic
  Instrument}}, \href{https://doi.org/10.3847/1538-3881/ad76a4}{\emph{\aj}
  {\bfseries 168} (2024) 245}.

\bibitem{DESI2022.KP1.Instr}
{DESI Collaboration}, B.~{Abareshi}, J.~{Aguilar}, S.~{Ahlen}, S.~{Alam},
  D.M.~{Alexander} et~al., \emph{{Overview of the Instrumentation for the Dark
  Energy Spectroscopic Instrument}},
  \href{https://doi.org/10.3847/1538-3881/ac882b}{\emph{\aj} {\bfseries 164}
  (2022) 207} [\href{https://arxiv.org/abs/2205.10939}{{\ttfamily
  2205.10939}}].

\bibitem{DESI2016a.Science}
{DESI Collaboration}, A.~{Aghamousa}, J.~{Aguilar}, S.~{Ahlen}, S.~{Alam},
  L.E.~{Allen} et~al., \emph{{The DESI Experiment Part I: Science,Targeting,
  and Survey Design}}, {\emph{arXiv e-prints} (2016) arXiv:1611.00036}
  [\href{https://arxiv.org/abs/1611.00036}{{\ttfamily 1611.00036}}].

\bibitem{Snowmass2013.Levi}
M.~{Levi}, C.~{Bebek}, T.~{Beers}, R.~{Blum}, R.~{Cahn}, D.~{Eisenstein}
  et~al., \emph{{The DESI Experiment, a whitepaper for Snowmass 2013}},
  {\emph{arXiv e-prints} (2013) arXiv:1308.0847}
  [\href{https://arxiv.org/abs/1308.0847}{{\ttfamily 1308.0847}}].

\bibitem{DESI2024.III.KP4}
{DESI Collaboration}, A.G.~Adame, J.~Aguilar, S.~Ahlen, S.~Alam, D.M.~Alexander
  et~al., \emph{{DESI 2024 III: Baryon Acoustic Oscillations from Galaxies and
  Quasars}}, \href{https://doi.org/10.48550/arXiv.2404.03000}{\emph{arXiv
  e-prints} (2024) arXiv:2404.03000}
  [\href{https://arxiv.org/abs/2404.03000}{{\ttfamily 2404.03000}}].

\bibitem{DESI2024.IV.KP6}
{DESI Collaboration}, A.G.~Adame, J.~Aguilar, S.~Ahlen, S.~Alam, D.M.~Alexander
  et~al., \emph{{DESI 2024 IV: Baryon Acoustic Oscillations from the Lyman
  Alpha Forest}}, \href{https://doi.org/10.48550/arXiv.2404.03001}{\emph{arXiv
  e-prints} (2024) arXiv:2404.03001}
  [\href{https://arxiv.org/abs/2404.03001}{{\ttfamily 2404.03001}}].

\bibitem{DESI2024.VI.KP7A}
{DESI Collaboration}, A.G.~Adame, J.~Aguilar, S.~Ahlen, S.~Alam, D.M.~Alexander
  et~al., \emph{{DESI 2024 VI: Cosmological Constraints from the Measurements
  of Baryon Acoustic Oscillations}},
  \href{https://doi.org/10.48550/arXiv.2404.03002}{\emph{arXiv e-prints} (2024)
  arXiv:2404.03002} [\href{https://arxiv.org/abs/2404.03002}{{\ttfamily
  2404.03002}}].

\bibitem{DESI:BAO-free}
E.A.~{Zaborowski}, P.~{Taylor}, K.~{Honscheid}, A.~{Cuceu}, A.~{de Mattia},
  D.~{Huterer} et~al., \emph{{A Sound Horizon-Free Measurement of $H_0$ in DESI
  2024}}, {\emph{arXiv e-prints} (2024) arXiv:2411.16677}
  [\href{https://arxiv.org/abs/2411.16677}{{\ttfamily 2411.16677}}].

\bibitem{Krolewski:2024jwj}
A.~Krolewski, W.J.~Percival and A.~Woodfinden, \emph{{A new method to determine
  $H_0$ from cosmological energy-density measurements}},
  \href{https://arxiv.org/abs/2403.19227}{{\ttfamily 2403.19227}}.

\bibitem{Brieden:2022heh}
S.~Brieden, H.~Gil-Mar\'\i{}n and L.~Verde, \emph{{A tale of two (or more)
  h's}}, \href{https://doi.org/10.1088/1475-7516/2023/04/023}{\emph{JCAP}
  {\bfseries 04} (2023) 023}
  [\href{https://arxiv.org/abs/2212.04522}{{\ttfamily 2212.04522}}].

\bibitem{Prada:2011uz}
F.~{Prada}, A.~{Klypin}, G.~{Yepes}, S.E.~{Nuza} and S.~{Gottloeber},
  \emph{{Measuring equality horizon with the zero-crossing of the galaxy
  correlation function}},
  \href{https://doi.org/10.48550/arXiv.1111.2889}{\emph{arXiv e-prints} (2011)
  arXiv:1111.2889} [\href{https://arxiv.org/abs/1111.2889}{{\ttfamily
  1111.2889}}].

\bibitem{DESI:ZeroCrossing}
Authors et~al., \emph{{Project: [383] Accurate Horizon-Size Measurement through
  Galaxy/QSO Correlation Zero-Crossing in DESI}}, {\emph{In prep.} }.

\bibitem{Eisenstein:EH98}
D.J.~{Eisenstein} and W.~{Hu}, \emph{{Baryonic Features in the Matter Transfer
  Function}}, \href{https://doi.org/10.1086/305424}{\emph{\apj} {\bfseries 496}
  (1998) 605} [\href{https://arxiv.org/abs/astro-ph/9709112}{{\ttfamily
  astro-ph/9709112}}].

\bibitem{Blas:class}
D.~{Blas}, J.~{Lesgourgues} and T.~{Tram}, \emph{{The Cosmic Linear Anisotropy
  Solving System (CLASS). Part II: Approximation schemes}},
  \href{https://doi.org/10.1088/1475-7516/2011/07/034}{\emph{\jcap} {\bfseries
  2011} (2011) 034} [\href{https://arxiv.org/abs/1104.2933}{{\ttfamily
  1104.2933}}].

\bibitem{Planck:params}
{Planck Collaboration}, N.~{Aghanim}, Y.~{Akrami}, M.~{Ashdown}, J.~{Aumont},
  C.~{Baccigalupi} et~al., \emph{{Planck 2018 results. VI. Cosmological
  parameters}}, \href{https://doi.org/10.1051/0004-6361/201833910}{\emph{\aap}
  {\bfseries 641} (2020) A6}
  [\href{https://arxiv.org/abs/1807.06209}{{\ttfamily 1807.06209}}].

\bibitem{Kaiser:RSD}
N.~{Kaiser}, \emph{{Clustering in real space and in redshift space}},
  \href{https://doi.org/10.1093/mnras/227.1.1}{\emph{\mnras} {\bfseries 227}
  (1987) 1}.

\bibitem{Bahr-Kalus:eBOSSTOerr}
B.~{Bahr-Kalus}, D.~{Parkinson} and E.-M.~{Mueller}, \emph{{Correction to:
  Measurement of the matter-radiation equality scale using the extended Baryon
  Oscillation Spectroscopic Survey Quasar Sample}},
  \href{https://doi.org/10.1093/mnras/stad3000}{\emph{\mnras} {\bfseries 526}
  (2023) 3248}.

\bibitem{Box:1964}
G.E.P.~Box and D.R.~Cox, \emph{An analysis of transformations},
  \href{https://doi.org/https://doi.org/10.1111/j.2517-6161.1964.tb00553.x}{\emph{Journal
  of the Royal Statistical Society: Series B (Methodological)} {\bfseries 26}
  (1964) 211}
  [\href{https://arxiv.org/abs/https://rss.onlinelibrary.wiley.com/doi/pdf/10.1111/j.2517-6161.1964.tb00553.x}{{\ttfamily
  https://rss.onlinelibrary.wiley.com/doi/pdf/10.1111/j.2517-6161.1964.tb00553.x}}].

\bibitem{Wang:distro}
M.S.~{Wang}, W.J.~{Percival}, S.~{Avila}, R.~{Crittenden} and D.~{Bianchi},
  \emph{{Cosmological inference from galaxy-clustering power spectrum:
  Gaussianization and covariance decomposition}},
  \href{https://doi.org/10.1093/mnras/stz829}{\emph{\mnras} {\bfseries 486}
  (2019) 951} [\href{https://arxiv.org/abs/1811.08155}{{\ttfamily
  1811.08155}}].

\bibitem{ChaussidonY1fnl}
E.~{Chaussidon}, C.~{Y{\`e}che}, A.~{de Mattia}, C.~{Payerne}, P.~{McDonald},
  A.J.~{Ross} et~al., \emph{{Constraining primordial non-Gaussianity with DESI
  2024 LRG and QSO samples}},
  \href{https://doi.org/10.48550/arXiv.2411.17623}{\emph{arXiv e-prints} (2024)
  arXiv:2411.17623} [\href{https://arxiv.org/abs/2411.17623}{{\ttfamily
  2411.17623}}].

\bibitem{ForemanMackey:emcee}
D.~{Foreman-Mackey}, D.W.~{Hogg}, D.~{Lang} and J.~{Goodman}, \emph{{emcee: The
  MCMC Hammer}}, \href{https://doi.org/10.1086/670067}{\emph{\pasp} {\bfseries
  125} (2013) 306} [\href{https://arxiv.org/abs/1202.3665}{{\ttfamily
  1202.3665}}].

\bibitem{DESI:targetting}
A.D.~{Myers}, J.~{Moustakas}, S.~{Bailey}, B.A.~{Weaver}, A.P.~{Cooper},
  J.E.~{Forero-Romero} et~al., \emph{{The Target-selection Pipeline for the
  Dark Energy Spectroscopic Instrument}},
  \href{https://doi.org/10.3847/1538-3881/aca5f9}{\emph{\aj} {\bfseries 165}
  (2023) 50} [\href{https://arxiv.org/abs/2208.08518}{{\ttfamily 2208.08518}}].

\bibitem{DESI:legacysurveyDR9}
D.~{Schlegel}, A.~{Dey}, D.~{Herrera}, S.~{Juneau}, M.~{Landriau}, D.~{Lang}
  et~al., \emph{{DESI Legacy Imaging Surveys Data Release 9}},  in
  \emph{American Astronomical Society Meeting Abstracts}, vol.~237 of
  \emph{American Astronomical Society Meeting Abstracts}, p.~235.03, Jan.,
  2021.

\bibitem{DESI:spectropipeline}
J.~{Guy}, S.~{Bailey}, A.~{Kremin}, S.~{Alam}, D.M.~{Alexander}, C.~{Allende
  Prieto} et~al., \emph{{The Spectroscopic Data Processing Pipeline for the
  Dark Energy Spectroscopic Instrument}},
  \href{https://doi.org/10.3847/1538-3881/acb212}{\emph{\aj} {\bfseries 165}
  (2023) 144} [\href{https://arxiv.org/abs/2209.14482}{{\ttfamily
  2209.14482}}].

\bibitem{BGS.TS.Hahn.2023}
C.~{Hahn}, M.J.~{Wilson}, O.~{Ruiz-Macias}, S.~{Cole}, D.H.~{Weinberg},
  J.~{Moustakas} et~al., \emph{{The DESI Bright Galaxy Survey: Final Target
  Selection, Design, and Validation}},
  \href{https://doi.org/10.3847/1538-3881/accff8}{\emph{\aj} {\bfseries 165}
  (2023) 253} [\href{https://arxiv.org/abs/2208.08512}{{\ttfamily
  2208.08512}}].

\bibitem{DESI:LRGs}
R.~{Zhou}, B.~{Dey}, J.A.~{Newman}, D.J.~{Eisenstein}, K.~{Dawson}, S.~{Bailey}
  et~al., \emph{{Target Selection and Validation of DESI Luminous Red
  Galaxies}}, \href{https://doi.org/10.3847/1538-3881/aca5fb}{\emph{\aj}
  {\bfseries 165} (2023) 58}
  [\href{https://arxiv.org/abs/2208.08515}{{\ttfamily 2208.08515}}].

\bibitem{DESI:ELGs}
A.~{Raichoor}, J.~{Moustakas}, J.A.~{Newman}, T.~{Karim}, S.~{Ahlen}, S.~{Alam}
  et~al., \emph{{Target Selection and Validation of DESI Emission Line
  Galaxies}}, \href{https://doi.org/10.3847/1538-3881/acb213}{\emph{\aj}
  {\bfseries 165} (2023) 126}
  [\href{https://arxiv.org/abs/2208.08513}{{\ttfamily 2208.08513}}].

\bibitem{DESI:QSOs}
E.~{Chaussidon}, C.~{Y{\`e}che}, N.~{Palanque-Delabrouille}, D.M.~{Alexander},
  J.~{Yang}, S.~{Ahlen} et~al., \emph{{Target Selection and Validation of DESI
  Quasars}}, \href{https://doi.org/10.3847/1538-4357/acb3c2}{\emph{\apj}
  {\bfseries 944} (2023) 107}
  [\href{https://arxiv.org/abs/2208.08511}{{\ttfamily 2208.08511}}].

\bibitem{RosadoMarin:DESIELGsyst}
A.~{Rosado-Marin} and {DESI Collaboration}, \emph{{Mitigating Imaging
  Systematics for DESI DR1 Emission Line Galaxies and Beyond}}, {\emph{in prep}
  (2024) }.

\bibitem{DESI:catalogues}
A.J.~{Ross}, J.~{Aguilar}, S.~{Ahlen}, S.~{Alam}, A.~{Anand}, S.~{Bailey}
  et~al., \emph{{The Construction of Large-scale Structure Catalogs for the
  Dark Energy Spectroscopic Instrument}},
  \href{https://doi.org/10.48550/arXiv.2405.16593}{\emph{arXiv e-prints} (2024)
  arXiv:2405.16593} [\href{https://arxiv.org/abs/2405.16593}{{\ttfamily
  2405.16593}}].

\bibitem{DESI2024.VII.KP7B}
{DESI Collaboration}, A.G.~{Adame}, J.~{Aguilar}, S.~{Ahlen}, S.~{Alam},
  D.M.~{Alexander} et~al., \emph{{DESI 2024 VII: Cosmological Constraints from
  the Full-Shape Modeling of Clustering Measurements}},
  \href{https://doi.org/10.48550/arXiv.2411.12022}{\emph{arXiv e-prints} (2024)
  arXiv:2411.12022} [\href{https://arxiv.org/abs/2411.12022}{{\ttfamily
  2411.12022}}].

\bibitem{Ross:eBOSScatalogue}
A.J.~{Ross}, J.~{Bautista}, R.~{Tojeiro}, S.~{Alam}, S.~{Bailey}, E.~{Burtin}
  et~al., \emph{{The Completed SDSS-IV extended Baryon Oscillation
  Spectroscopic Survey: Large-scale structure catalogues for cosmological
  analysis}}, \href{https://doi.org/10.1093/mnras/staa2416}{\emph{\mnras}
  {\bfseries 498} (2020) 2354}
  [\href{https://arxiv.org/abs/2007.09000}{{\ttfamily 2007.09000}}].

\bibitem{Feldman:FKPweights}
H.A.~{Feldman}, N.~{Kaiser} and J.A.~{Peacock}, \emph{{Power-Spectrum Analysis
  of Three-dimensional Redshift Surveys}},
  \href{https://doi.org/10.1086/174036}{\emph{\apj} {\bfseries 426} (1994) 23}
  [\href{https://arxiv.org/abs/astro-ph/9304022}{{\ttfamily
  astro-ph/9304022}}].

\bibitem{Ross:DESICat}
A.J.~{Ross}, J.~{Aguilar}, S.~{Ahlen}, S.~{Alam}, A.~{Anand}, S.~{Bailey}
  et~al., \emph{{The Construction of Large-scale Structure Catalogs for the
  Dark Energy Spectroscopic Instrument}},
  \href{https://doi.org/10.48550/arXiv.2405.16593}{\emph{arXiv e-prints} (2024)
  arXiv:2405.16593} [\href{https://arxiv.org/abs/2405.16593}{{\ttfamily
  2405.16593}}].

\bibitem{Chaussidon:regressis_and_DESIQSOsel}
E.~{Chaussidon}, C.~{Y{\`e}che}, N.~{Palanque-Delabrouille}, A.~{de Mattia},
  A.D.~{Myers}, M.~{Rezaie} et~al., \emph{{Angular clustering properties of the
  DESI QSO target selection using DR9 Legacy Imaging Surveys}},
  \href{https://doi.org/10.1093/mnras/stab3252}{\emph{\mnras} {\bfseries 509}
  (2022) 3904} [\href{https://arxiv.org/abs/2108.03640}{{\ttfamily
  2108.03640}}].

\bibitem{Chaussidon:blinding}
E.~{Chaussidon}, A.~{de Mattia}, C.~{Y{\`e}che}, J.~{Aguilar}, S.~{Ahlen},
  D.~{Brooks} et~al., \emph{{Blinding scheme for the scale-dependence bias
  signature of local primordial non-Gaussianity for DESI 2024}},
  \href{https://doi.org/10.48550/arXiv.2406.00191}{\emph{arXiv e-prints} (2024)
  arXiv:2406.00191} [\href{https://arxiv.org/abs/2406.00191}{{\ttfamily
  2406.00191}}].

\bibitem{Andrade:blinding}
U.~{Andrade}, J.~{Mena-Fern{\'a}ndez}, H.~{Awan}, A.J.~{Ross}, S.~{Brieden},
  J.~{Pan} et~al., \emph{{Validating the Galaxy and Quasar Catalog-Level
  Blinding Scheme for the DESI 2024 analysis}},
  \href{https://doi.org/10.48550/arXiv.2404.07282}{\emph{arXiv e-prints} (2024)
  arXiv:2404.07282} [\href{https://arxiv.org/abs/2404.07282}{{\ttfamily
  2404.07282}}].

\bibitem{Yamamoto:estimator}
K.~{Yamamoto}, M.~{Nakamichi}, A.~{Kamino}, B.A.~{Bassett} and H.~{Nishioka},
  \emph{{A Measurement of the Quadrupole Power Spectrum in the Clustering of
  the 2dF QSO Survey}},
  \href{https://doi.org/10.1093/pasj/58.1.93}{\emph{\pasj} {\bfseries 58}
  (2006) 93} [\href{https://arxiv.org/abs/astro-ph/0505115}{{\ttfamily
  astro-ph/0505115}}].

\bibitem{Hand:pypower}
N.~{Hand}, Y.~{Li}, Z.~{Slepian} and U.~{Seljak}, \emph{{An optimal FFT-based
  anisotropic power spectrum estimator}},
  \href{https://doi.org/10.1088/1475-7516/2017/07/002}{\emph{\jcap} {\bfseries
  2017} (2017) 002} [\href{https://arxiv.org/abs/1704.02357}{{\ttfamily
  1704.02357}}].

\bibitem{Chuang:EZmocks}
C.-H.~{Chuang}, F.-S.~{Kitaura}, F.~{Prada}, C.~{Zhao} and G.~{Yepes},
  \emph{{EZmocks: extending the Zel'dovich approximation to generate mock
  galaxy catalogues with accurate clustering statistics}},
  \href{https://doi.org/10.1093/mnras/stu2301}{\emph{\mnras} {\bfseries 446}
  (2015) 2621} [\href{https://arxiv.org/abs/1409.1124}{{\ttfamily 1409.1124}}].

\bibitem{Maksimova:AbacusSummit}
N.A.~{Maksimova}, L.H.~{Garrison}, D.J.~{Eisenstein}, B.~{Hadzhiyska},
  S.~{Bose} and T.P.~{Satterthwaite}, \emph{{ABACUSSUMMIT: a massive set of
  high-accuracy, high-resolution N-body simulations}},
  \href{https://doi.org/10.1093/mnras/stab2484}{\emph{\mnras} {\bfseries 508}
  (2021) 4017} [\href{https://arxiv.org/abs/2110.11398}{{\ttfamily
  2110.11398}}].

\bibitem{Garrison:Abacus}
L.H.~{Garrison}, D.J.~{Eisenstein}, D.~{Ferrer}, N.A.~{Maksimova} and
  P.A.~{Pinto}, \emph{{The ABACUS cosmological N-body code}},
  \href{https://doi.org/10.1093/mnras/stab2482}{\emph{\mnras} {\bfseries 508}
  (2021) 575} [\href{https://arxiv.org/abs/2110.11392}{{\ttfamily
  2110.11392}}].

\bibitem{Hartlap:factor}
J.~{Hartlap}, P.~{Simon} and P.~{Schneider}, \emph{{Why your model parameter
  confidences might be too optimistic. Unbiased estimation of the inverse
  covariance matrix}},
  \href{https://doi.org/10.1051/0004-6361:20066170}{\emph{\aap} {\bfseries 464}
  (2007) 399} [\href{https://arxiv.org/abs/astro-ph/0608064}{{\ttfamily
  astro-ph/0608064}}].

\bibitem{Percival:covariancefactor}
W.J.~{Percival}, A.J.~{Ross}, A.G.~{S{\'a}nchez}, L.~{Samushia}, A.~{Burden},
  R.~{Crittenden} et~al., \emph{{The clustering of Galaxies in the SDSS-III
  Baryon Oscillation Spectroscopic Survey: including covariance matrix
  errors}}, \href{https://doi.org/10.1093/mnras/stu112}{\emph{\mnras}
  {\bfseries 439} (2014) 2531}
  [\href{https://arxiv.org/abs/1312.4841}{{\ttfamily 1312.4841}}].

\bibitem{KP4s7-Rashkovetskyi}
M.~Rashkovetskyi, D.~Forero-S{\'a}nchez, A.~de~Mattia, D.J.~Eisenstein,
  N.~Padmanabhan, H.~Seo et~al., \emph{{Semi-analytical covariance matrices for
  two-point correlation function for DESI 2024 data}},
  \href{https://doi.org/10.48550/arXiv.2404.03007}{\emph{arXiv e-prints} (2024)
  arXiv:2404.03007} [\href{https://arxiv.org/abs/2404.03007}{{\ttfamily
  2404.03007}}].

\bibitem{Beutler:window}
F.~Beutler and P.~McDonald, \emph{{Unified galaxy power spectrum measurements
  from 6dFGS, BOSS, and eBOSS}},
  \href{https://doi.org/10.1088/1475-7516/2021/11/031}{\emph{JCAP} {\bfseries
  11} (2021) 031} [\href{https://arxiv.org/abs/2106.06324}{{\ttfamily
  2106.06324}}].

\bibitem{deMattia:radintconst}
A.~{de Mattia} and V.~{Ruhlmann-Kleider}, \emph{{Integral constraints in
  spectroscopic surveys}},
  \href{https://doi.org/10.1088/1475-7516/2019/08/036}{\emph{\jcap} {\bfseries
  2019} (2019) 036} [\href{https://arxiv.org/abs/1904.08851}{{\ttfamily
  1904.08851}}].

\bibitem{Planck:png}
{Planck Collaboration}, Y.~{Akrami}, F.~{Arroja}, M.~{Ashdown}, J.~{Aumont},
  C.~{Baccigalupi} et~al., \emph{{Planck 2018 results. IX. Constraints on
  primordial non-Gaussianity}},
  \href{https://doi.org/10.1051/0004-6361/201935891}{\emph{\aap} {\bfseries
  641} (2020) A9} [\href{https://arxiv.org/abs/1905.05697}{{\ttfamily
  1905.05697}}].

\bibitem{Cunnington:HITO}
S.~{Cunnington}, \emph{{Detecting the power spectrum turnover with H I
  intensity mapping}},
  \href{https://doi.org/10.1093/mnras/stac576}{\emph{\mnras} {\bfseries 512}
  (2022) 2408} [\href{https://arxiv.org/abs/2202.13828}{{\ttfamily
  2202.13828}}].

\bibitem{Brieden:ShapeFit}
S.~{Brieden}, H.~{Gil-Mar{\'\i}n} and L.~{Verde}, \emph{{ShapeFit: extracting
  the power spectrum shape information in galaxy surveys beyond BAO and RSD}},
  \href{https://doi.org/10.1088/1475-7516/2021/12/054}{\emph{\jcap} {\bfseries
  2021} (2021) 054} [\href{https://arxiv.org/abs/2106.07641}{{\ttfamily
  2106.07641}}].

\bibitem{DESI2024.V.KP5}
{DESI Collaboration}, A.G.~{Adame}, J.~{Aguilar}, S.~{Ahlen}, S.~{Alam},
  D.M.~{Alexander} et~al., \emph{{DESI 2024 V: Full-Shape Galaxy Clustering
  from Galaxies and Quasars}},
  \href{https://doi.org/10.48550/arXiv.2411.12021}{\emph{arXiv e-prints} (2024)
  arXiv:2411.12021} [\href{https://arxiv.org/abs/2411.12021}{{\ttfamily
  2411.12021}}].

\bibitem{Fixsen}
D.J.~{Fixsen}, \emph{{The Temperature of the Cosmic Microwave Background}},
  \href{https://doi.org/10.1088/0004-637X/707/2/916}{\emph{\apj} {\bfseries
  707} (2009) 916} [\href{https://arxiv.org/abs/0911.1955}{{\ttfamily
  0911.1955}}].

\bibitem{Pantheon:data}
D.~{Scolnic}, D.~{Brout}, A.~{Carr}, A.G.~{Riess}, T.M.~{Davis}, A.~{Dwomoh}
  et~al., \emph{{The Pantheon+ Analysis: The Full Data Set and Light-curve
  Release}}, \href{https://doi.org/10.3847/1538-4357/ac8b7a}{\emph{\apj}
  {\bfseries 938} (2022) 113}
  [\href{https://arxiv.org/abs/2112.03863}{{\ttfamily 2112.03863}}].

\bibitem{Pantheon:params}
D.~{Brout}, D.~{Scolnic}, B.~{Popovic}, A.G.~{Riess}, A.~{Carr}, J.~{Zuntz}
  et~al., \emph{{The Pantheon+ Analysis: Cosmological Constraints}},
  \href{https://doi.org/10.3847/1538-4357/ac8e04}{\emph{\apj} {\bfseries 938}
  (2022) 110} [\href{https://arxiv.org/abs/2202.04077}{{\ttfamily
  2202.04077}}].

\bibitem{Union:3}
D.~{Rubin}, G.~{Aldering}, M.~{Betoule}, A.~{Fruchter}, X.~{Huang}, A.G.~{Kim}
  et~al., \emph{{Union Through UNITY: Cosmology with 2,000 SNe Using a Unified
  Bayesian Framework}},
  \href{https://doi.org/10.48550/arXiv.2311.12098}{\emph{arXiv e-prints} (2023)
  arXiv:2311.12098} [\href{https://arxiv.org/abs/2311.12098}{{\ttfamily
  2311.12098}}].

\bibitem{DES:SN5YR}
M.~{Vincenzi}, D.~{Brout}, P.~{Armstrong}, B.~{Popovic}, G.~{Taylor},
  M.~{Acevedo} et~al., \emph{{The Dark Energy Survey Supernova Program:
  Cosmological Analysis and Systematic Uncertainties}},
  \href{https://doi.org/10.3847/1538-4357/ad5e6c}{\emph{\apj} {\bfseries 975}
  (2024) 86} [\href{https://arxiv.org/abs/2401.02945}{{\ttfamily 2401.02945}}].

\bibitem{Riess:SH0ES}
A.G.~{Riess}, W.~{Yuan}, L.M.~{Macri}, D.~{Scolnic}, D.~{Brout}, S.~{Casertano}
  et~al., \emph{{A Comprehensive Measurement of the Local Value of the Hubble
  Constant with 1 km s$^{-1}$ Mpc$^{-1}$ Uncertainty from the Hubble Space
  Telescope and the SH0ES Team}},
  \href{https://doi.org/10.3847/2041-8213/ac5c5b}{\emph{\apjl} {\bfseries 934}
  (2022) L7} [\href{https://arxiv.org/abs/2112.04510}{{\ttfamily 2112.04510}}].

\bibitem{Chevallier:CPL}
M.~{Chevallier} and D.~{Polarski}, \emph{{Accelerating Universes with Scaling
  Dark Matter}},
  \href{https://doi.org/10.1142/S0218271801000822}{\emph{International Journal
  of Modern Physics D} {\bfseries 10} (2001) 213}
  [\href{https://arxiv.org/abs/gr-qc/0009008}{{\ttfamily gr-qc/0009008}}].

\bibitem{Linder:CPL}
E.V.~{Linder}, \emph{{Exploring the Expansion History of the Universe}},
  \href{https://doi.org/10.1103/PhysRevLett.90.091301}{\emph{\prl} {\bfseries
  90} (2003) 091301} [\href{https://arxiv.org/abs/astro-ph/0208512}{{\ttfamily
  astro-ph/0208512}}].

\bibitem{bahr_kalus_2025_17018002}
B.~Bahr-Kalus, D.~Parkinson and K.~Lodha, \emph{Data for "model-independent
  measurement of the matter-radiation equality scale in desi 2024"},  Sept.,
  2025.
\newblock 10.5281/zenodo.17018002.

\bibitem{2025arXiv250314745D}
{DESI Collaboration}, M.~{Abdul-Karim}, A.G.~{Adame}, D.~{Aguado},
  J.~{Aguilar}, S.~{Ahlen} et~al., \emph{{Data Release 1 of the Dark Energy
  Spectroscopic Instrument}},
  \href{https://doi.org/10.48550/arXiv.2503.14745}{\emph{arXiv e-prints} (2025)
  arXiv:2503.14745} [\href{https://arxiv.org/abs/2503.14745}{{\ttfamily
  2503.14745}}].

\bibitem{2020ASPC..527..307T}
G.~{Taffoni}, U.~{Becciani}, B.~{Garilli}, G.~{Maggio}, F.~{Pasian}, G.~{Umana}
  et~al., \emph{{CHIPP: INAF Pilot Project for HTC, HPC and HPDA}},  in
  \emph{Astronomical Data Analysis Software and Systems XXIX}, R.~{Pizzo},
  E.R.~{Deul}, J.D.~{Mol}, J.~{de Plaa} and H.~{Verkouter}, eds., vol.~527 of
  \emph{Astronomical Society of the Pacific Conference Series}, p.~307, Jan.,
  2020, \href{https://doi.org/10.48550/arXiv.2002.01283}{DOI}
  [\href{https://arxiv.org/abs/2002.01283}{{\ttfamily 2002.01283}}].

\bibitem{2020ASPC..527..303B}
S.~{Bertocco}, D.~{Goz}, L.~{Tornatore}, A.~{Ragagnin}, G.~{Maggio},
  F.~{Gasparo} et~al., \emph{{INAF Trieste Astronomical Observatory Information
  Technology Framework}},  in \emph{Astronomical Data Analysis Software and
  Systems XXIX}, R.~{Pizzo}, E.R.~{Deul}, J.D.~{Mol}, J.~{de Plaa} and
  H.~{Verkouter}, eds., vol.~527 of \emph{Astronomical Society of the Pacific
  Conference Series}, p.~303, Jan., 2020,
  \href{https://doi.org/10.48550/arXiv.1912.05340}{DOI}
  [\href{https://arxiv.org/abs/1912.05340}{{\ttfamily 1912.05340}}].

\end{thebibliography}\endgroup

%% or
%% [B] Manual formatting (see below)
%% (i) We suggest to always provide author, title and journal data or doi:
%% in short all the informations that clearly identify a document.
%% (ii) please avoid comments such as "For a review", "For some examples",
%% "and references therein" or move them in the text. In general, please leave only references in the bibliography and move all
%% accessory text in footnotes.
%% (iii) Also, please have only one work for each \bibitem.

% \begin{thebibliography}{99}

% \bibitem{a}
% Author,
% \emph{Title},
% \emph{J. Abbrev.} {\bf vol} (year) pg.

% \bibitem{b}
% Author,
% \emph{Title},
% arxiv:1234.5678.

% \bibitem{c}
% Author,
% \emph{Title},
% Publisher (year).

% \end{thebibliography}

\end{document}